# Traceable localization enables accurate integration of quantum emitters and photonic structures with high yield

Craig R. Copeland,[1] Adam L. Pintar,[2] Ronald G. Dixson,[1] Ashish Chanana,[1] Kartik Srinivasan,[1,3] Daron A. Westly,[1] B. Robert Ilic,[1,4] Marcelo I. Davanco,[1] and Samuel M. Stavis[1,*]

[1]Microsystems and Nanotechnology Division, [2]Statistical Engineering Division, [4]CNST NanoFab, National Institute of Standards and Technology, Gaithersburg, Maryland 20899, USA

[3]Joint Quantum Institute, NIST/University of Maryland, College Park, Maryland 20742, USA

[*]samuel.stavis@nist.gov

Traceability to the International System of Units (SI) is fundamental to measurement accuracy and reliability. In this study, we demonstrate subnanometer traceability of localization microscopy, establishing a metrological foundation for the maturation and application of super-resolution methods. To do so, we create a master standard by measuring the positions of submicrometer apertures in an array by traceable atomic-force microscopy. We perform correlative measurements of this master standard by optical microscopy, calibrating scale factor and correcting aberration effects. We introduce the concept of a localization uncertainty field due to optical localization errors and scale factor uncertainty, with regions of position traceability to within a 68 % coverage interval of ± 1 nm. These results enable localization metrology with high throughput, which we apply to measure working standards that we fabricate by electron-beam lithography, validating the accuracy of mean pitch and closing the loop for disseminating and integrating reference arrays. We then apply our novel methods to calibrate an optical microscope with a sample cryostat, accounting for thermal contraction by use of a submicrometer pillar array in silicon as a reference material and elucidating complex distortion. This new calibration enables the accurate integration of quantum emitters and photonic structures with high yield, as we demonstrate theoretically through simulations of the dependence of the Purcell factor of radiative enhancement on registration errors across a wide field. Our study illuminates the challenges and opportunities of achieving traceable localization for comparison of position data across lithography and microscopy systems, from ambient to cryogenic temperatures.

## INTRODUCTION

Localization microscopy has left the diffraction limit of a few hundred nanometers in the rearview optics, enabling nanoscale measurements in diverse applications ranging from biological imaging to photonic integration[1-3]. As novel methods mature and reproducibility concerns deepen[4,5], a better understanding of localization uncertainty becomes increasingly important. Whereas uncertainty due to the random effect of shot noise can range down to the length scale of one nanometer and below[6-8], systematic effects can be orders of magnitude larger and vary unpredictably across the imaging field[8]. Identification and correction of such effects requires comprehensive calibration of optical microscopes[8-11], which is uncommon, leading to a common discrepancy of precision and accuracy, and potential overconfidence in localization data. Moreover, for localization microscopy, no previous study has established a continuous chain of calibrations with corresponding uncertainties that is reliably traceable to the International System of Units (SI). The traceability of localization data is much more than a formality, and is indeed a requirement to ensure accuracy. This fundamental issue becomes critical when reliable position data and corresponding sample dimensions are necessary for meaningful comparison and accurate registration across systems and studies.

A calibration can be only as good as the standard providing a reference. Unofficial standards for applications of localization microscopy to biological imaging include fluorescent particles for calibration of point spread functions and registration of localization data at different wavelengths[10,12,13], molecular nanostructures for calibration of local scale factors[14,15], and nanoscale apertures for all of these calibrations, as well as global calibrations of the imaging field and stability tests[8,16]. Of these unofficial standards, aperture arrays are relatively uncommon but feature flexibility of design from the top down and use under different imaging conditions, accessibility for correlative microscopy to establish traceability, and experimental stability and reusability[8,16]. In contrast, lithographic standards are more common in applications of localization microscopy to the integration of quantum emitters and photonic structures, often involving electron-beam lithography, which presents natural opportunities to place alignment markers for calibration of scale factors and correction of aberration effects[17-20].

In a previous study, we fabricated aperture arrays by electron-beam lithography and tested aperture placement[8]. Two lithography systems each used two interferometers to control stage positions and correct for electron-optical aberrations within the patterning process. By localizing apertures and comparing placements by the two systems, we estimated a mean distance between apertures that differed by one part in five thousand, or approximately 1 nm, and random placement errors of approximately 2 nm. Although the implication was placement accuracy at the nanometer scale, these test results were insufficient to claim traceability. Moreover, the significant difference of mean distance raises questions about the variability of our lithography process to set critical dimensions for demanding applications.

Even with a standard in hand or on chip, at least four fundamental challenges impede the traceability of localization microscopy. The first challenge is matching all experimental conditions—system optics, imaging modes, sample positions, and localization analyses—between calibration and experiment[8]. Any inconsistency can degrade accuracy and compromise reliability, such as by limiting the applicability of calibration data that diverge from the experimental context of localization microscopy[15,21,22]. This issue compounds the second challenge of calibrating the scale factor, or mean magnification, or image pixel size of an optical microscope with low uncertainty. These limits also pertain to the third challenge of sampling the imaging field with nanostructures that are suitable for localization microscopy and that probe field nonuniformity at the scale from one to ten wavelengths[8,10]. The fourth challenge is to maintain the integrity of the calibration chain by quantitating uncertainty and validating results. No previous study has met all of these challenges. We do so in the present study, establishing a traceability chain for localization microscopy, achieving subnanometer uncertainty across a wide field, closing the loop of making and measuring standards, and extending our methods to cryogenic temperatures[23-25]. These results establish a foundation of traceability for cryogenic applications including correlative electron and photon microscopy for biological imaging[5,26-31], potentially with better precision at lower temperatures[24,25] but typically without supporting accuracy, as well as integration of quantum emitters with photonic structures with high yield by various microscopy methods[17-20,32-36], as we investigate further (Figure 1).

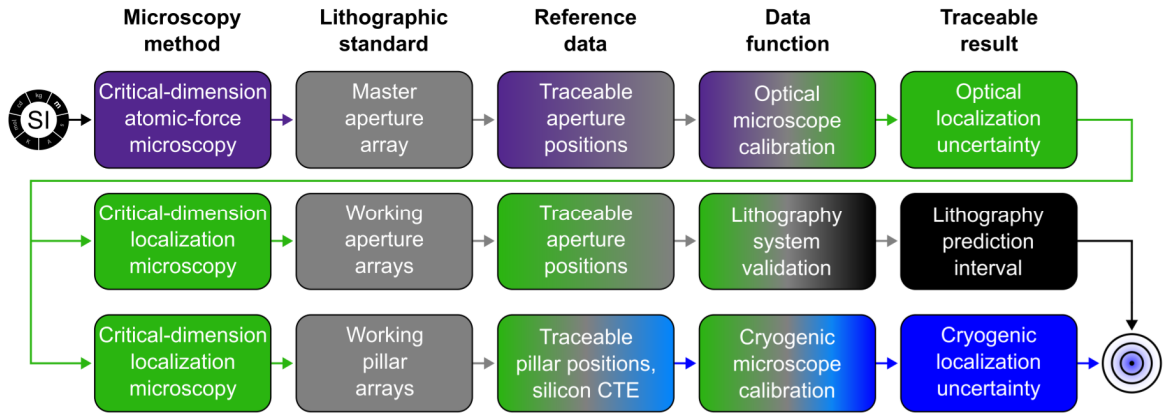

**Figure 1**. Overview. A process workflow and traceability chain show links extending from SI units of nanometers to the accurate integration of a quantum dot and a photonic microresonator in the form of a bullseye target. Purple denotes critical-dimension atomic-force microscopy. Gray denotes array structures prior to microscopy measurements that characterize critical dimensions. Green denotes optical microscopy at ambient temperature. Black denotes electron-beam lithography after validation, with the potential to set critical dimensions without microscopy measurements. Light blue denotes reference data for coefficient of thermal expansion (CTE). Dark blue denotes optical microscopy at cryogenic temperatures.

In summary of our study, we begin by creating a master standard. We measure the positions of an aperture array by critical-dimension atomic-force microscopy[37,38], which is traceable to the SI through interferometric calibrations and transfer standards[39,40]. Correlative measurements of this master standard by optical microscopy lead to a novel calibration of scale factor and optimal correction of aberration effects[8]. Statistical models of measurement variability lead to the new concept of an uncertainty field in localization microscopy, shifting the paradigm not only from precision to accuracy[8] but further to widefield traceability. The resulting capability of critical-dimension localization microscopy enables traceable characterization of working standards with high throughput. A statistical meta-analysis validates lithographic accuracy, closing the loop of producing reference arrays for dissemination and integration into processes and devices. We apply all of these concepts to a comprehensive calibration of a widefield microscope with a sample cryostat and custom optics, enabling traceable localization at cryogenic temperatures for accurate integration of quantum emitters and photonic structures with high yield. This important application presents specific demands for accuracy across lithography and microscopy systems[17-20]. Potential errors result from multiple sources that are nonobvious at the state of the art, including lithographic and cryogenic variation of reference dimensions to set scale factors, and complex distortion from custom optics. The resulting errors tend to increase with field extent, so that a comprehensive calibration is critical for leveraging the throughput and scalability of widefield imaging. Representative simulations show the effect of these errors on Purcell factor, which is a key performance metric for radiative enhancement upon integrating quantum emitters into microresonator structures. Our study illuminates the challenges and opportunities of achieving traceability in localization microscopy, and establishes a metrological foundation for widefield scalability and yield engineering in the integration of quantum emitters and photonic structures.

**Table 1. Terms and symbols**

| Term | Symbol |
|---|---|
| **General distance analysis** | |
| True distance between two points | $\Delta$ |
| Experimental measurement of distance between two points by AFM or OM | $D^{AFM}, D^{OM}$ |
| Experimental measurement of distance between two points in the x or y direction by OM | $D_x^{OM}, D_y^{OM}$ |
| Uncertainty of $D^{OM}$ | $u_{D^{OM}}$ |
| Uncertainty of $D_x^{OM}$ or $D_y^{OM}$ between one localization result and a reference position | $u_{D_x^{OM}}, u_{D_y^{OM}}$ |
| Uncertainty of $D_x^{OM}$ or $D_y^{OM}$ between two localization results | $u_{D_x^{OM}}^{II}, u_{D_y^{OM}}^{II}$ |
| Scale factor for OM | $S$ |
| Scale factor uncertainty for OM | $\sigma_S$ |
| Scale factor error for OM | $\epsilon_S$ |
| **Aperture pair analysis** | |
| Aperture pair index | $i$ |
| Replicate measurement index | $j$ |
| Number of replicate measurements of aperture pair distance by AFM | $n_{D^{AFM}}$ |
| Experimental measurement of aperture pair distance by AFM | $D_{ij}^{AFM} = \Delta_i + d_{ij}^{AFM} + \delta_i^{AFM}$ |
| Random error of distance that is observable between replicates for AFM | $d_{ij}^{AFM}$ |
| Variance of $d_{ij}^{AFM}$ assuming the same for each aperture pair and replicate, and dividing by $n_{D^{AFM}}$ | $\sigma_{d^{AFM}}^2 / n_{D^{AFM}}$ |
| Random error of distance that is unobservable between replicates for AFM | $\delta_i^{AFM}$ |
| Variance of $\delta_i^{AFM}$ assuming the same for each aperture pair | $\sigma_{\delta^{AFM}}^2$ |
| Variance of $D^{AFM}$ | $\sigma_{D^{AFM}}^2 = \sigma_{\delta^{AFM}}^2 + \sigma_{d^{AFM}}^2 / n_{D^{AFM}}$ |
| Number of replicate measurements of aperture pair distance by OM | $n_{D^{OM}}$ |
| Experimental measurement of aperture pair distance by OM | $D_{ij}^{OM} = \Delta_i + d_{ij}^{OM} + \delta_i^{OM}$ |
| Random error of distance that is observable between replicates for OM | $d_{ij}^{OM}$ |
| Variance of $d_{ij}^{OM}$ assuming the same for each aperture pair and replicate, and dividing by $n_{D^{OM}}$ | $\sigma_{d^{OM}}^2 / n_{D^{OM}}$ |
| Random error of distance that is unobservable between replicates for OM | $\delta_i^{OM}$ |
| Variance of $\delta_i^{OM}$ assuming the same for each aperture pair | $\sigma_{\delta^{OM}}^2$ |
| Variance of $D^{OM}$ | $\sigma_{D^{OM}}^2 = \sigma_{\delta^{OM}}^2 + \sigma_{d^{OM}}^2 / n_{D^{OM}}$ |
| Distance deviation after averaging over replicates | $D_{i\cdot}^{OM} - D_{i\cdot}^{AFM}$ |
| Variance of distance deviations over aperture pairs | $\sigma_{D^{OM} - D^{AFM}}^2$ |

AFM is atomic-force microscopy. OM is optical microscopy.
For clarity, we include only symbols that appear in this study. For example, $D_x^{AFM}$ does not appear.
For completeness, $u_{D_x^{OM}} = u_{D_x^{OM}}^I$ and $u_{D_y^{OM}} = u_{D_y^{OM}}^I$, which we revisit in Supplementary Note S3. For clarity, we simplify this notation in the text.
A dot symbol for $j$ in a subscript denotes an average over replicate measurements.
Additional information and discussions of these quantities are in Supplementary Table S3, Supplementary Note S2, and Supplementary Note S3.

**RESULTS AND DISCUSSION**

We begin with one of the aperture arrays from our previous test[8], for which we had designed a pitch of 5000 nm and found a deviation of approximately 1 nm. We presently image 21 pairs of adjacent apertures in triplicate, and one pair in duplicate, by critical-dimension atomic-force microscopy (Figure 2a-b) (Table 1) (Supplementary Table S1). The two axes of the atomic-force microscope scan independently, with each axis probing the aperture sidewalls with a flared tip, measuring the distance between 11 different pairs of apertures with a resolution of less than 0.1 nm and a relative uncertainty of approximately one part in ten thousand[37-40]. We report all uncertainties as 68 % coverage intervals (Supplementary Note S1[41-45]) (Supplementary Table S2). This relative uncertainty of $10^{-4}$ results from calibration of the mean scale factor, which involves correction of scanfield distortion of the atomic-force microscope (Figure 3a). We select for analysis sidewall positions ranging from 30 nm to 95 nm above the zero plane of the silica substrate (Supplementary Figure S1), assuming that this nearly vertical region determines the transmission of light through apertures in subsequent optical microscopy. A comparison of elliptical models and centroid analyses of the sidewall positions to localize each aperture yields consistent results. Least-squares fits of elliptical models with uniform weighting smooth the scatter of the sidewall positions, reducing the pooled standard deviation of the distance between pairs of adjacent apertures from 1.50 nm to 0.98 nm, so we proceed with this analysis.

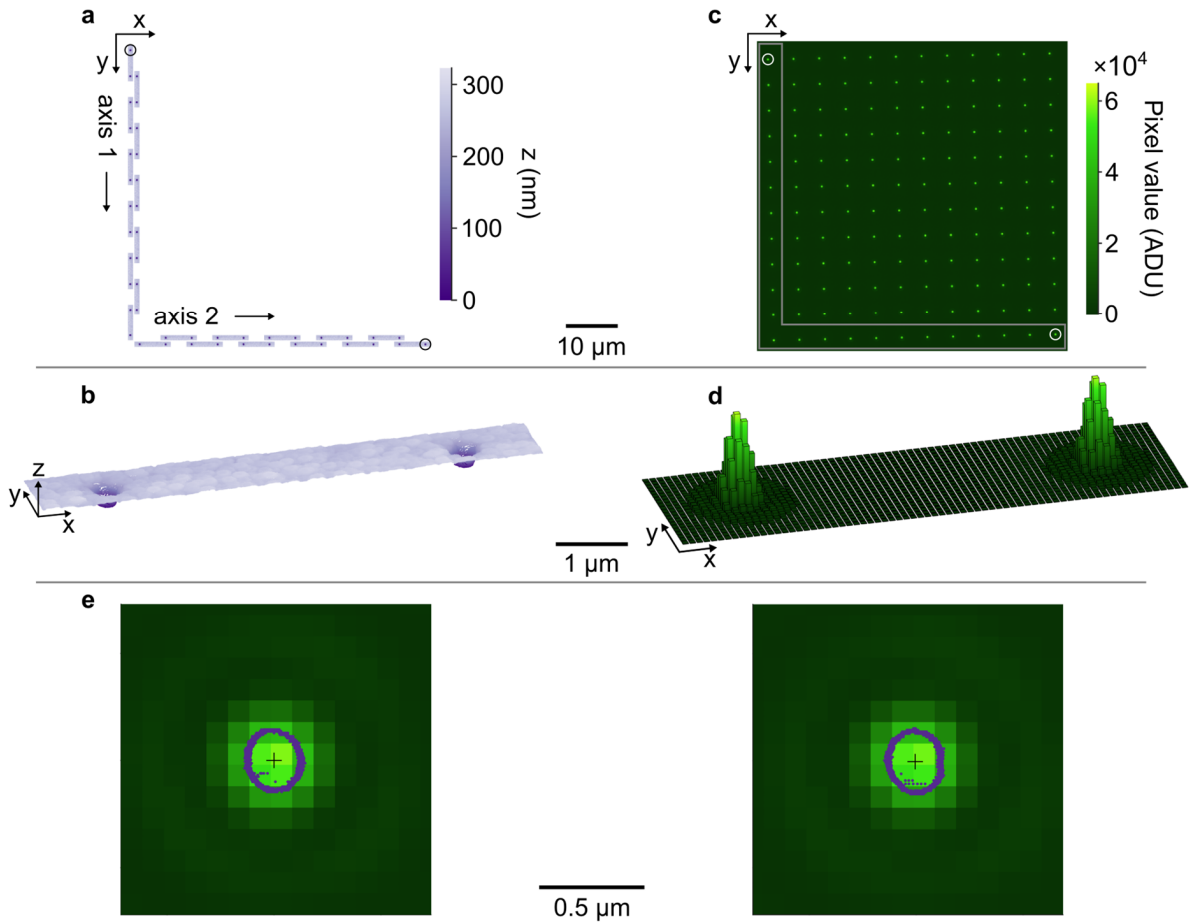

**Figure 2.** Correlative microscopy. (**a-b**) Atomic-force micrographs showing (a) separate images of the 22 pairs of apertures comprising the left column and bottom row of apertures in the array, and (b) an image of a representative pair of apertures. (**c-d**) Optical micrographs showing (c) the entire aperture array and (d) an image of the same representative pair of apertures. The gray outline in (c) indicates the same apertures as in (a). Circles in (a, c) indicate two corner apertures. (**e**) Correlative micrographs showing the direct overlay of image data from the two microscopy methods for the aperture pair in (b-d). Black crosses mark the positions resulting from localization analyses. Position uncertainties are smaller than the cross linewidths.

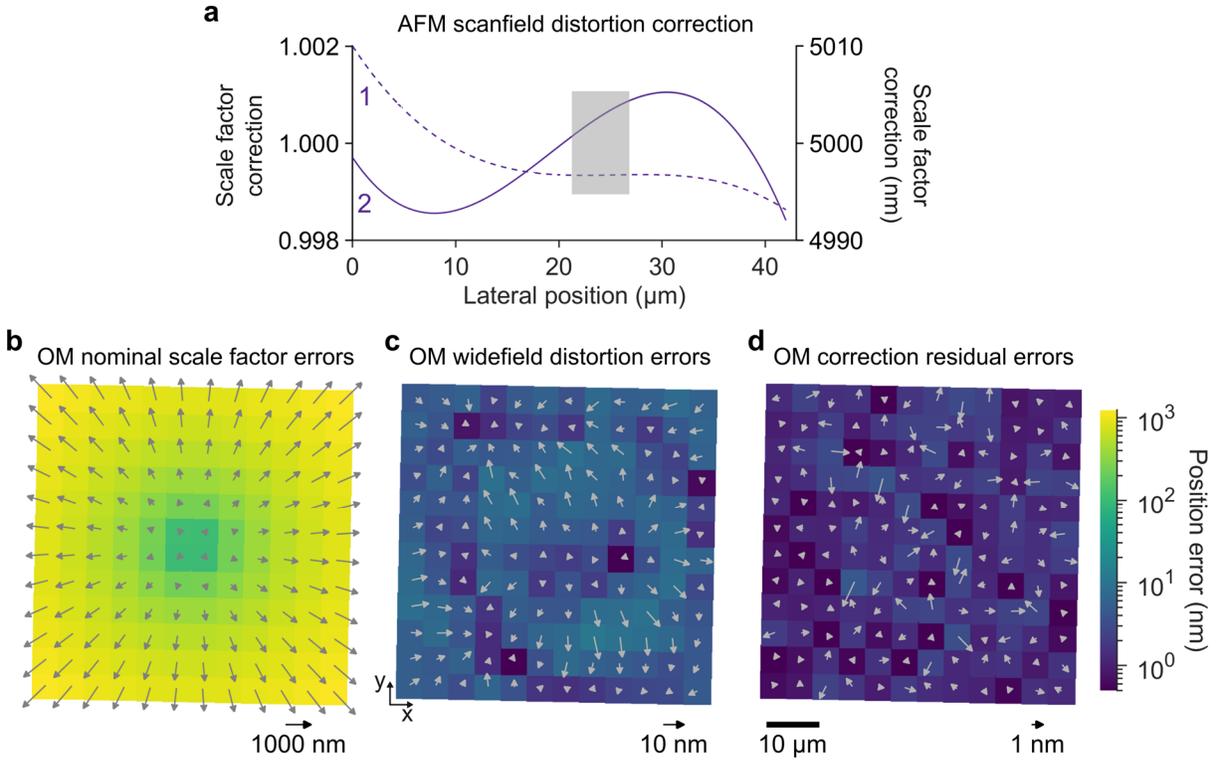

**Figure 3.** Microscopy calibrations and corrections. (**a**) Plot showing scanfield distortion corrections for (dash line) axis 1 and (solid line) axis 2 of the atomic-force microscope. The gray box indicates the region of interest. (**b-d**) Vector plots and logarithmic color maps showing position errors for optical microscopy, (b) assuming the nominal scale factor or magnification of the system is accurate, (c) after calibration of mean scale factor, and (d) after scale factor calibration and widefield distortion correction. The optical calibration is insensitive to the evident rotation of the aperture array relative to the coordinate system of the imaging sensor.

Complementary statistical models of two types enable analyses of the aperture pair distances, leading to a master standard (Supplementary Note S2) (Table 1). Fixed-effect models for both atomic-force microscopy and optical microscopy assess localization accuracy and enable interstudy comparison. An autoregressive model explicitly accounts for the correlation of adjacent aperture pairs in atomic-force microscopy (Figure 2a), but the comparable results of the two types of models are similar. Accordingly, the fixed-effect model for atomic-force microscopy yields a mean distance of 5000.72 nm ± 0.24 nm for axis 1, 5000.69 nm ± 0.06 nm for axis 2, and 5000.71 nm ± 0.13 nm for both axes. These uncertainties account for variability from replicate measurements of the 22 pairs of apertures (Supplementary Table S1). The propagation of scale factor uncertainty (Supplementary Table S2) results in a traceable mean distance of 5000.71 nm $\pm$ 0.54 nm. Although this pitch of the master standard is near to the nominal value, subsequent tests of more standards are necessary to sample variation of the lithographic process, with important implications for setting critical dimensions.

We record optical micrographs of the aperture array using our previous methods[8] and modify the following calibration to achieve traceability. We image the entire array by optical microscopy near best focus (Figure 2c-d) with 1000 replicates (Supplementary Table S1). A similarity transformation between an ideal array of positions and the apparent positions resulting from localization analysis determines the mean magnification of the optical microscope, setting the scale factor in the form of an image pixel size. To achieve traceability, the mean distance between aperture pairs of the master standard from atomic-force microscopy, rather than the nominal pitch of the array from electron-beam lithography, defines the pitch of the ideal array. This calibration reveals that assuming the nominal magnification of the objective lens results in a scale factor error of 3.1 %[8], causing egregious position errors (Figure 3b). We revisit the effects of calibrating the scale factor using an erroneous reference dimension. Position errors remain after the similarity transformation (Figure 3c), due to widefield distortion and other aberration effects, which a Zernike polynomial model then corrects[8] (Figure 3d). In another modification from our previous work, the correlation of aperture distances between optical and atomic-force microscopy allows a traceable optimization of the number of polynomial terms in the correction model. For 144 apertures in a field of 3600 µm$^2$, a Zernike model that includes all polynomials up to a Noll index of 41 for x and 48 for y minimizes root-mean-square deviations of distance between the same aperture pairs by the two microscopy methods (Figure 3c-d). Importantly, both scale factor errors and distortion errors can be even more significant over larger imaging fields[8].

A comparison of aperture pair distances from the two microscopy methods manifests uncertainty components that are rich with information (Table 1) (Supplementary Note S2). The aperture pair distances are in evident agreement (Figure 4) with deviations that have a mean of zero within uncertainty (Supplementary Table S1) and variances that enable determination of limiting uncertainty components for optical microscopy. In the fixed-effect models, the total variances are (Eq. 1) $\sigma^2_{D^{AFM}} = \sigma^2_{\delta^{AFM}} + \sigma^2_{d^{AFM}}/n_{D^{AFM}}$ for atomic-force microscopy and (Eq. 2) $\sigma^2_{D^{OM}} = \sigma^2_{\delta^{OM}} + \sigma^2_{d^{OM}}/n_{D^{OM}}$ for optical microscopy (Supplementary Note S2). We divide the variances from replicate measurements, $\sigma^2_{d^{OM}}$ and $\sigma^2_{d^{AFM}}$, by the sample sizes $n_{D^{AFM}}$ and $n_{D^{OM}}$, as we average distance values over replicates. The variances $\sigma^2_{\delta^{AFM}}$ and $\sigma^2_{\delta^{OM}}$ are from localization errors, such as from non-uniform scale or deviations of localization models from aperture images. For optical microscopy, $\sigma_{d^{OM}}/\sqrt{2}$ is the empirical localization precision and $\sigma^2_{\delta^{OM}}/2$ is the variance of position from errors that are unobservable in temporal replicates[8] (Supplementary Table S3). Assuming the independence of random effects, the sample variance of the distance deviations is the sum of Eq. 1 and Eq. 2, (Eq. 3) $\sigma^2_{D^{OM}-D^{AFM}} = \sigma^2_{\delta^{OM}} + \sigma^2_{d^{OM}}/n_{D^{OM}} + \sigma^2_{\delta^{AFM}} + \sigma^2_{d^{AFM}}/n_{D^{AFM}}$. Pooling the sample variances of the replicate measurements over all aperture pairs yields values for $\sigma^2_{d^{AFM}}$ and $\sigma^2_{d^{OM}}$, separately for each axis of the atomic-force microscope, due to the higher variability of axis 1 from a control algorithm to improve sidewall tracking (Figure 4). We estimate $\sigma^2_{\delta^{AFM}}$ for each axis. For axis 1, the scale factor is nearly constant in the region of interest (Figure 3a), yielding a negligible value of $\sigma_{\delta^{AFM}}$ (Table 2). For axis 2, the scale factor varies nearly linearly in the region of interest (Figure 3a), and we use a scale factor correction from near the mid-point of the aperture pairs. The repeatability of sample positioning is within 0.2 µm, yielding a larger but still negligible value of $\sigma_{\delta AFM}$ for this axis (Table 2). For this reason, replication effects dominate the variance of atomic-force microscopy for both axes. For optical microscopy, averaging $n_{D^{OM}} = 1000$ replicates, with $0.6 \times 10^6$ signal photons per image, reduces $\sigma_{d^{OM}}^2/n_{D^{OM}}$ to a negligible value (Supplementary Table S4)[46]. Having estimates of all other terms in Eq. 3, we solve for $\sigma_{\delta^{OM}}^2$ (Table 2) (Supplementary Table S4).

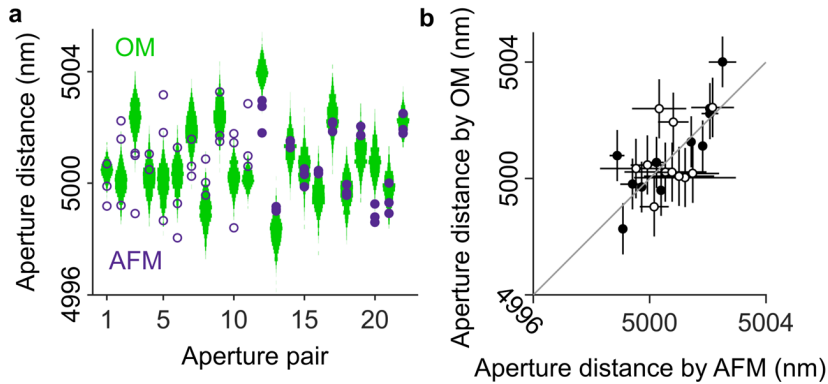

**Figure 4**. Aperture distance measurements. (**a**) Plot showing the correlation of aperture distances from (purple circles, hollow circles are axis 1, solid circles are axis 2) atomic-force microscopy (AFM) and (green violin histograms) optical microscopy (OM). (**b**) Plot showing the correlation of distances for the different measurements, with a reduction of the data in (a) to mean values and 68 % coverage intervals in (b).

**Table 2. Distance uncertainty evaluation**

| Uncertainty component | Evaluation | Absolute value for x direction (nm) | Absolute value for y direction (nm) |
|---|---|---|---|
| $\sigma_{D^{OM}-D^{AFM}}$ | Type A, measurement | 0.88 ± 0.21 | 1.10 ± 0.26 |
| $\sigma_{d^{AFM}}/\sqrt{n_{D^{AFM}}}$ | Type A, measurement | 0.19 ± 0.05 | 0.79 ± 0.19 |
| $\sigma_{\delta^{AFM}}$ | Type B, estimate | 0.02 | 0.004 |
| $\sigma_{d^{OM}}/\sqrt{n_{D^{OM}}}$ | Type A, measurement | 0.017 ± 0.004 | 0.017 ± 0.004 |
| $\sigma_{\delta^{OM}}$ | Type A, Eq. (3) | 0.86 ± 0.21 | 0.77 ± 0.32 |
| $\sigma_{\delta^{OM}}$ | Type A, Reference 8 | 0.88 ± 0.28 | 1.02 ± 0.27 |

The solution for $\sigma_{\delta^{OM}}$ depends on a Type B evaluation of $\sigma_{\delta^{AFM}}$, which would typically result in an overall categorization of a Type B evaluation for $\sigma_{\delta^{OM}}$. However, the effect of $\sigma_{\delta^{AFM}}$ is negligible, so that the solution for $\sigma_{\delta^{OM}}$ effectively involves only Type A evaluations of the three other components, resulting in an effective overall categorization of Type A.

The resulting estimate of optical localization error enables an interstudy validation and provides new insight into imaging field nonuniformity. The quantities of $\sigma_{\delta^{OM}}$, in both the $x$ and $y$ directions, agree within uncertainty with those from our previous study with a larger field of 40,000 µm² (Table 2)[8], providing a consistency check for this optical microscope and further indicating that its limiting localization errors are spatially random and independent of field area. The effect of fabrication precision, or actual deviations of the aperture positions from the nominal design, on the Zernike model causes localization errors of less than 0.1 nm[8]. Wavefront errors that cause nonuniformity of the imaging field over micrometer length scales are a probable cause of the remaining localization errors. The following calibration of a cryogenic microscope supports this likelihood, showing qualitatively similar but quantitatively larger localization errors.

Optical localization error sets a lower bound of traceable position uncertainty, but scale factor uncertainty becomes increasingly important across a wide field. This effect limits the accurate integration of quantum emitters and photonic structures, among other applications requiring reliable registration of position data, such as correlative microscopy. To quantify this trend, $\sigma_{D^{OM}}$ sums in quadrature with a scale factor uncertainty component. For the position uncertainty of a single point from localization analysis, relative to a reference point that requires no localization analysis, such as a pixel position, the values of $\sigma_{D^{OM}}$ reduce by a factor of $\sqrt{2}$. The uncertainty is then, (Eq. 4) $u_{D_x^{OM}} \approx \sqrt{\left(\sigma_{D_x^{OM}}/\sqrt{2}\right)^2 + \left(D_x^{OM} \times \sigma_S\right)^2}$ (Figure 5) (Supplementary Table S2) (Table 2), where $D_x^{OM}$ is the distance in the $x$ direction from a reference point, with an analogous expression for the $y$ direction, and $\sigma_S$ is the relative uncertainty of scale factor. A relative error of scale factor, $\epsilon_S$, can exceed $\sigma_S$ and yield a similar effect of greater magnitude[8] (Figure 3c). A general expression for the uncertainty of distance $D^{OM}$, including localization uncertainty for two points, is in Supplementary Note S3. The first term in Eq. 4 is constant, whereas the second term scales with distance (Figure 5a). This trend is characteristic of coordinate-measuring machines, closing the gap between such systems and conventional optical microscopes. Equation 4 and its analogue in the y direction describe an uncertainty field with two regions of positions with subnanometer traceability (Figure 5b). In the inner region, the maximum uncertainty is less than ± 1.0 nm across an area of 180 µm². In the outer region, the mean uncertainty is less than ± 1.0 nm across an area of 390 µm². The uncertainty field exhibits asymmetric variation around the center due to different values of $\sigma_{\delta_x^{OM}}$ and $\sigma_{\delta_y^{OM}}$ (Figure 5), clarifying both the limiting components and spatial extent of subnanometer uncertainty[7,8]. These results emphasize the challenge of calibrating scale factor and propagating its uncertainty in localization microscopy, which are uncommon.

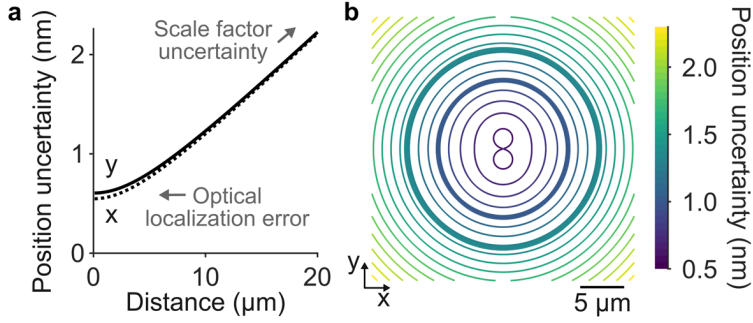

**Figure 5.** Traceable position uncertainty. (**a**) Plot showing position uncertainty of optical microscopy $u_{D_{x,y}^{OM}}$ as a function of distance $D_{x,y}^{OM}$ from a reference point for the (solid) $x$ and (dash) $y$ directions. (**b**) Contour plot showing the corresponding uncertainty field for position relative to the field center. Two bold contours are limits of (inner) maximum and (outer) mean uncertainty of less than ± 1.0 nm. Contour intervals are 0.1 nm.

In a final test of the two microscopy methods, we measure the diagonal distance between two corner apertures (Figure 2a, c), neglecting the common uncertainty of scale factor to isolate other components of uncertainty. For atomic-force microscopy, assuming that the array axes are orthogonal, and that off-axis effects of fabrication precision are negligible, summation of distances between the intermediate 22 aperture pairs (Figure 2a) yields a diagonal distance of 77792.76 nm ± 1.94 nm, with random effects dominating the uncertainty. For optical microscopy, direct localization of only the two corner apertures (Figure 2c) yields a diagonal distance of 77792.36 nm ± 0.82 nm, with systematic effects dominating the uncertainty. Thus, the critical dimensions are in subnanometer agreement, validating axis orthogonality for atomic-force microscopy and the accuracy of localization analysis for optical microscopy. Moreover, these results show that optical localization yields half the uncertainty from components that are not common to both microscopy methods, and higher throughput by a factor of $10^5$.

This high throughput yields a new capability for fabrication process metrology, enabling critical-dimension localization microscopy of five aperture arrays and three silicon pillar arrays as working standards (Supplementary Table S5). The silicon pillar arrays have advantageous material properties as cryogenic microscopy standards in the following application. A statistical meta-analysis[44] yields a consensus mean pitch of 4999.80 nm ± 0.98 nm, an estimate of pitch variability from fabrication corresponding to a standard deviation of 2.70 nm, and a 68 % prediction interval for the fabrication of additional working standards ranging from 4997.42 nm to 5002.13 nm (Supplementary Note S4). These traceable results validate the subnanometer accuracy of mean pitch and, due to the variation of process parameters, provide a conservative estimate of the reliability of producing replicate standards by electron-beam lithography. This result is an important step toward establishing statistical process control for fabricating and integrating working standards into processes and devices, potentially without additional characterization. However, process variability may exceed scale uncertainty, and does so by a factor of five in the present study. Therefore, characterizing each working standard to ensure traceability, rather than assuming that the critical dimension of a working standard is equal to a nominal value or conforms to a manufacturer specification, is necessary in the following application.

In an application that exploits all of our methods and results (Figure 1), we demonstrate a novel calibration of a cryogenic localization microscope – an optical microscope with the sample and objective lens inside of a cryostat, and custom optics outside of the cryostat (Supplementary Note S5). Beyond the general interest in performing localization and correlative microscopy at cryogenic temperatures, the integration of quantum emitters into photonic structures provides specific demands for localization traceability. At the state of the art, epitaxial growth yields the self-assembly of quantum dots that radiate single photons with fast lifetime, high efficiency, and high coherence. These useful properties place quantum dots at the forefront of light sources for information systems. However, current growth processes result in quantum dots with random positions in semiconductor substrates. Previous studies have used localization microscopy to measure these lateral positions at cryogenic temperatures[17-20], guiding the subsequent placement of photonic resonators by electron-beam lithography at ambient temperatures. This integration process requires the accurate registration of position data across instruments and temperatures, with optimal coupling of emitters and resonators occurring within a registration error of a few tens of nanometers.[17,32] Accordingly, reducing registration errors is a topic of emerging interest, with previous studies having reduced errors due to barrel distortion[32], reduced aberration effects by a perspective transformation[19], and improved localization of alignment markers by eliminating potential effects of laser misalignment[47]. As well, a review noted the potential utility of our methods for a distortion correction.[33] Presently, we study not only the optical aberrations of cryogenic microscopes but also the thermal deviations of reference dimensions, and we reveal pitfalls of sparsely sampling the imaging field to achieve the traceable calibration of such a measurement system. Building on this calibration, we combine our microscopy and lithography results to introduce a comprehensive model of the effects of registration errors on the Purcell factor and corresponding yield that is theoretically achievable through the integration process.

To create novel working standards that are suitable for the imaging mode of this microscope system, and that have reliable reference data for coefficient of thermal expansion,[48] we pattern submicrometer pillars in a silicon (100) substrate by electron-beam lithography at approximately 293 K, yielding micropillar arrays with a nominal pitch of 5000 nm (Supplementary Figure S3, Supplementary Note S5). Elemental silicon is an excellent reference material for this purpose. Whereas a fused silica or borosilicate substrate, such as a microscope coverslip, would present questions about the type and purity of the glass[49], silicon (100) wafers are widely available and readily amenable to standard lithographic processes. We characterize the pitch of each array by critical-dimension localization microscopy at approximately 293 K, finding a mean pitch across three arrays of 5001.71 nm ± 0.54 nm (Supplementary Note S5, Supplementary Table S6). Using reference data for silicon (100) from a study involving cryogenic interferometry,[48] we estimate a net contraction of pitch of 0.021509 % ± 0.000003 %, or 1.07580 nm ± 0.00015 nm, upon cooling to approximately 1.8 K. Importantly, other semiconductor materials that are relevant for photonic integration contract more and have reference data that are less certain. For example, gallium arsenide contracts by nearly four times this amount[50], presenting both challenges and opportunities that we revisit. For our purposes, the coefficient of thermal expansion for silicon (100) is negligible below approximately 20 K, so we neglect any temperature deviations in this range.

The novel calibration of a cryogenic microscope reveals sources of error that are more complex and less evident than those of a conventional microscope, illuminating potential dark uncertainty[44] in previous studies. To study these effects, we load the reference arrays into the cryostat and record optical micrographs at a peak wavelength of 940 nm near best focus at sample temperatures of approximately 293 K and 1.8 K. We localize the pillars and calibrate the imaging field at each temperature (Supplementary Note S5), yielding image pixel sizes (Supplementary Figure S4) and maps of position errors (Figure 6, Supplementary Figure S5).

Scale factor errors can be significant even for conventional microscopes[8] (Figure 3b), and our study reveals latent sources of such errors for microscope systems with sample cryostats and custom optics. Custom imaging systems may lack a nominal scale factor. Accordingly, previous studies[17-20] have used alignment markers of a frame type fabricated by electron-beam lithography for a cryogenic calibration of magnification by a measurement of the distance between two reference positions that span the imaging field. In addition to the unreliable assumption of the accuracy of the lithography system to set the reference distance, we find that such measurements yield only a sparse sampling of complex distortion effects among other aberration effects that can vary at the scale of one to ten wavelengths. These effects can lead to large errors of scale factor (Supplementary Figure S6a), elucidating a nonobvious but critical problem with a typical two-point calibration of scale factor. Moreover, previous disregard of the contraction of gallium arsenide substrates at cryogenic temperatures[50] results in scale factor errors that severely limit the accuracy of these studies (Supplementary Figure S6b). These various sources of error can have significant impact on the efficiency of photonic integration, as we show subsequently.

Separately from the calibration of mean magnification, we observe position errors resulting from complex distortion (Figure 6, Supplementary Figure S5). The resulting position errors differ from those of previous studies involving conventional microscopes[8,21], implying the contributions of both objective lenses and custom optics, and emphasizing the utility of our methods to identify and correct systematic effects. An additional microscope with an alternate placement of an objective lens within a sample cryostat[20] shows even larger errors (Supplementary Figure S7), indicating the general need for calibrating such systems. Without correction, complex distortion causes position errors as large as 170 nm, with root-mean-square values of 50 nm in x and 55 nm in y for a representative pillar array, whereas the empirical localization precision is less than 2.0 nm (Supplementary Table S7). Our correction reduces these errors by more than a factor of three, and the improvement can be larger for different microscopes (Supplementary Figure S7) and wider fields[8,10]. We separately measure some components of the total position errors and estimate others to quantitate localization errors of 14.2 nm in x and 13.9 nm in y (Figure 6, Supplementary Note S5, Supplementary Table S7). In applying the calibration of image pixel size and distortion correction to experimental data, additional sources of error increase uncertainty. These include random effects such as shot noise, as well as systematic effects such as variation of axial[8] and lateral position, which alter apparent position errors, as we discuss subsequently. These additional sources of error become evident upon applying the calibration from images of one array to those of a different array (Supplementary Table S8).

Our results show a distinct hierarchy of optical localization error and scale factor uncertainty. For the cryogenic microscope with custom optics, which is more susceptible than a conventional microscope to aberration effects from potential imperfections and misalignments of optical components, localization error is the dominant source of uncertainty for the entire imaging field that we study, with a maximum uncertainty of position due to scale factor uncertainty of less than 6 nm. This result is due to significant lateral variation of localization errors at the micrometer scale, which vary with lower spatial frequency in our conventional microscope, and which Zernike models can only partially capture for the cryogenic microscope. In contrast, interpolant models can fully capture such errors at the pillar locations, but sampling still limits efficacy, resulting in little improvement over Zernike models (Supplementary Note S5, Supplementary Table S8). While future work can improve upon these results with even denser sampling of the imaging field, the present calibration significantly improves localization accuracy across a wide field and enables a dramatic improvement of process yield, as follows.

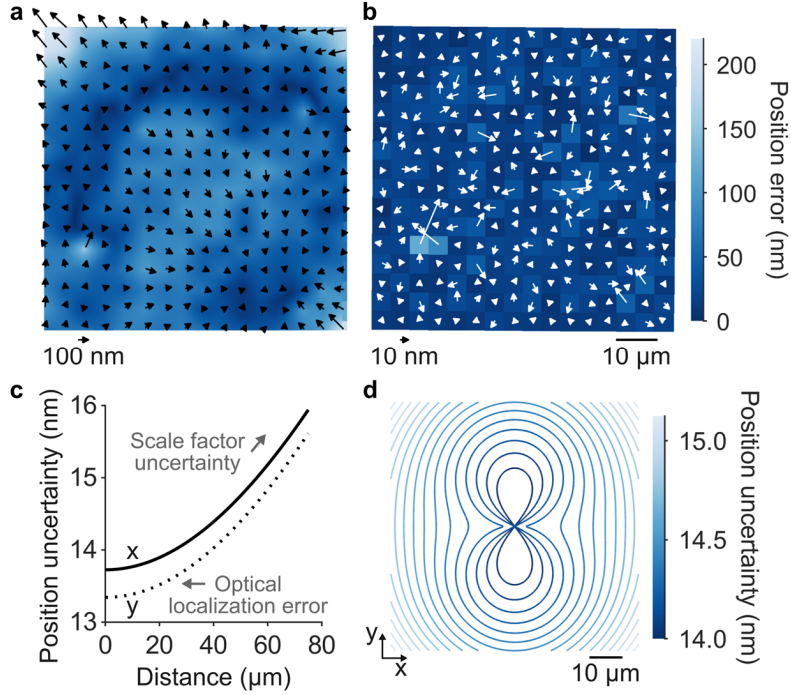

**Figure 6.** Cryogenic localization calibration. (**a-b**) Vector plots and color maps showing position errors (a) before and (b) after correction for imaging at a temperature of approximately 1.8 K. (**c**) Plot showing traceable position uncertainty as a function of distance from a reference point for the (solid) $x$ and (dash) $y$ directions. (**d**) Contour plot showing the corresponding uncertainty field for position in relative to the field center.

To assess the theoretical yield of integrating quantum emitters with photonic structures, we introduce a comprehensive model of registration errors between quantum dots and photonic microresonators[17-20], combining errors and uncertainties that result from calibration, measurement, and fabrication (Supplementary Note S6). We then consider Purcell factor as a performance metric that quantifies the radiative rate enhancement of a quantum dot[51], depending on electric field strength, which varies considerably over tens of nanometers in typical microresonator geometries. To obtain a representative dependence of Purcell factor on registration error, we interpolate the simulation results from Supplementary Figure 6e of Reference 17 for a dipole emitter within a bullseye target cavity, with a wavelength of 948.02 nm and azimuthal angles of 0 rad, π/4 rad, and π/2 rad[17] (Supplementary Figure S8). Our model results in Purcell factor histograms for each position in the imaging field (Supplementary Figure S9), which reduce to mean values (Figure 7). The practical limitations of the calibrations that we describe in the previous sections, the potential divergence of reflection and scattering from reference structures and emission from quantum dots[47] at slightly different wavelengths[17-20], and any deviation of the Purcell factor dependence between simulations and experiments[17] all motivate future work to test and build on our foundational model.

Our comprehensive model enables a systematic study of the theoretical impact of microscope calibration and device fabrication on Purcell factor through sequential reductions of registration error. Beginning with the state of the art[17-20] (Figure 7, Scenario 1), we reduce registration errors from localization measurements, consistent with implementation of our calibration methods, defining five process scenarios. We then reduce registration errors from the process of fabricating photonic structures to elucidate the effects, although such a reduction in fabrication error is purely theoretical in this study, defining three additional process scenarios. A tabular summary of these process scenarios is in Figure 7a, representative plots of mean Purcell factor across the imaging field are in Figure 7b, and histograms of mean Purcell factor for each process scenario are in Figure 7c. Further details of errors and uncertainties, probability distributions, and data reduction are in Supplementary Note S6, Supplementary Table S9, and Supplementary Figure S9.

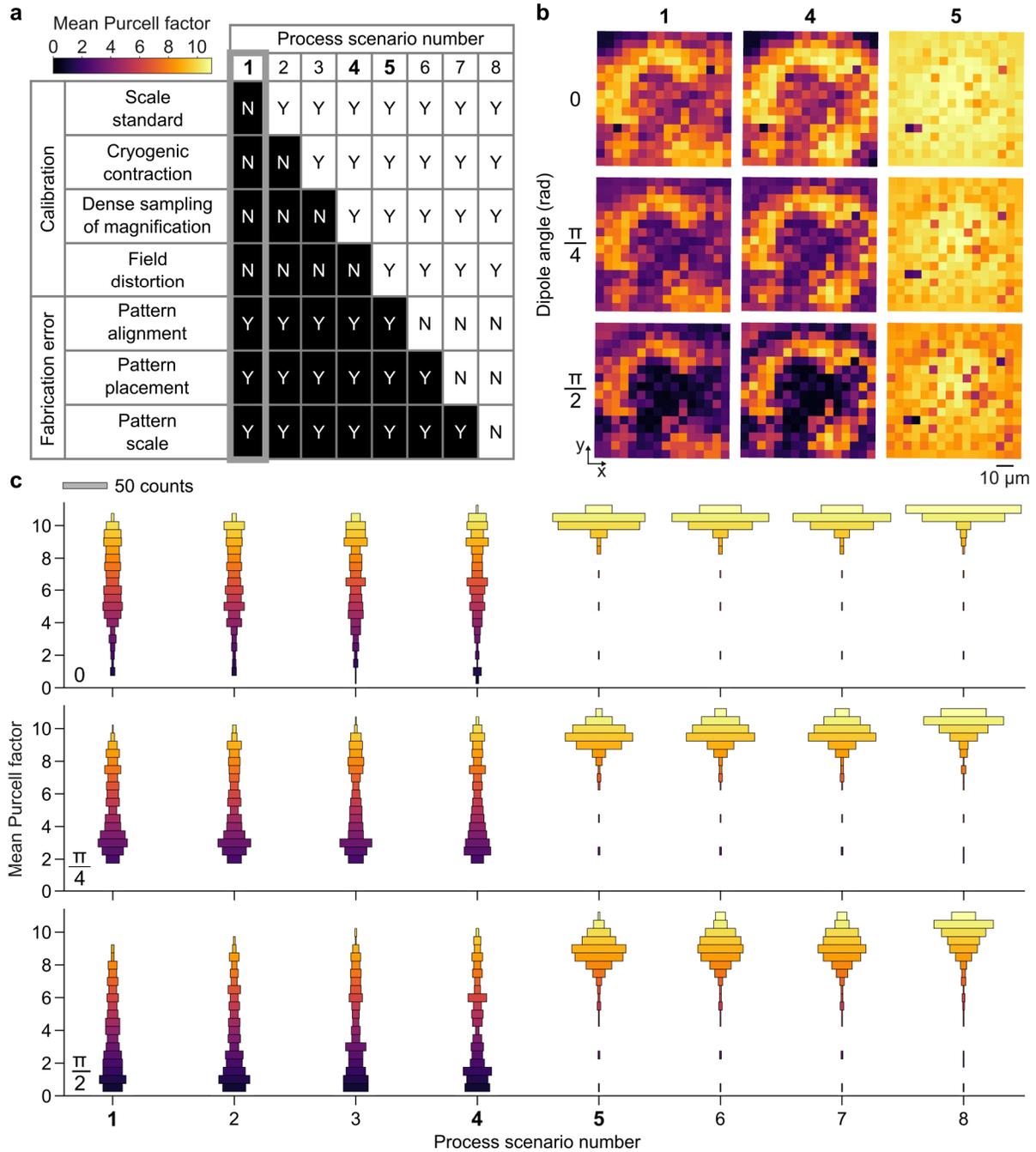

**Figure 7.** Purcell factor improvement due to accurate integration. (**a**) Table showing combinations of four calibrations and three sources of fabrication error comprising eight process scenarios. The color map at top left applies to panels (b-c). (**b**) Plots showing mean Purcell factor across the imaging field for Scenarios 1, 4, and 5, and dipole angles of (top) 0 rad, (middle) π/4 rad, and (bottom) π/2 rad. (**c**) Plots showing histograms of mean Purcell factor for each process scenario and dipole angles of (top) 0 rad, (middle) π/4 rad, and (bottom) π/2 rad.

The state of the art corresponds in some approximation to Scenario 1, prior to comprehensive calibration of the cryogenic localization microscope and potential improvement of the lithographic fabrication process (Figure 7a). The result is a broad distribution of Purcell factors, varying across the imaging field (Figure 7b) and ranging from the minimum to the maximum value (Figure 7c). Although this result allows for the fabrication of a few devices that demonstrate high performance, the low yield leaves room for improvement toward a reliable and scalable process, as follows.

The effects of microscope calibration are evident in Scenarios 2 through 5. For setting the scale factor, these effects include calibrating a standard rather than assuming its nominal dimensions (Figure 7, Scenario 2), accounting for the net contraction of the standard at cryogenic temperatures (Figure 7, Scenario 3), and using an array standard that densely samples an imaging field with complex distortion, rather than using alignment marks that sparsely sample the imaging field (Figure 7, Scenario 4). These calibrations provide significant but minor improvements in Purcell factor relative to the major effect of distortion (Figure 7). However, a larger imaging field is not only possible but also desirable to increase the throughput of device fabrication and characterization, and would result in larger effects of scale factor errors. Moreover, other photonic structures require registration accuracy at the scale of 1 nm[52], making these scale factor calibrations potentially critical. The final calibration step is the distortion correction, greatly improving both the magnitude and uniformity of Purcell factor near its maximum value across the imaging field (Figure 7, Scenario 5). Maps of Purcell factor for three scenarios—the state of the art (Scenario 1), improvement by comprehensive calibration of scale factor (Scenario 4), and further improvement by distortion correction (Scenario 5) – summarize these results (Figure 7b).

The effects of registration errors from fabrication of photonic structures by electron-beam lithography are evident in Scenarios 6 through 8. These include errors from overlay or alignment, precision of feature placement, and scale factor error. Removing these sources of fabrication error produces meaningful increases of mean Purcell factor, as Figure 7c shows, motivating future work. These results also indicate that the deterministic fabrication of quantum emitters will likely involve lithographic systems that yield imperfect placement of emitters, requiring emitter localization after fabrication for optimum integration[53-55]. After characterization, emitter arrays could eliminate any potential divergence of measurement conditions between calibration and experiment.

Completing the extraction of information from our model, we explore the theoretical yield as a function of process scenario number and Purcell factor threshold (Figure 8, Supplementary Figure S10). The results show an overall improvement of theoretical yield throughout the process scenarios, with the largest effect due to the distortion correction. Small decreases in yield for process scenarios 1 through 4 can occur for lower values of Purcell factor and are due to a partial cancellation of scale factor and complex distortion errors. Such an interaction of the two major effects that we calibrate is interesting, difficult to predict, and an unreliable approach to improve yield. Depending on the dipole angle and Purcell factor threshold, the improvement beyond the state of the art from comprehensive calibration ranges from one to two orders of magnitude (Figure 8, Supplementary Figure S10, Scenarios 1 and 5). The distortion correction allows for a high yield that remains nearly constant as the Purcell factor threshold increases to 8, whereas at the state of the art the yield decreases through this threshold range (Supplementary Figure S10). Interestingly, achieving high yield for the highest Purcell factor thresholds is challenging even after comprehensive calibration of the optical microscope and theoretical removal of errors from fabrication (Figure 8, Supplementary Figure S10, Scenario 8). This limit corresponds to the variability and resulting uncertainty of scale factor in our calibration of the cryogenic microscope (Supplementary Table S6). This effect is likely due to the variability of sample positioning between calibrations, with contributions from localization errors, and future work may address these issues. Importantly, these results highlight the importance of a comprehensive calibration and establish a foundation for future studies to improve process yield and optimize device performance.

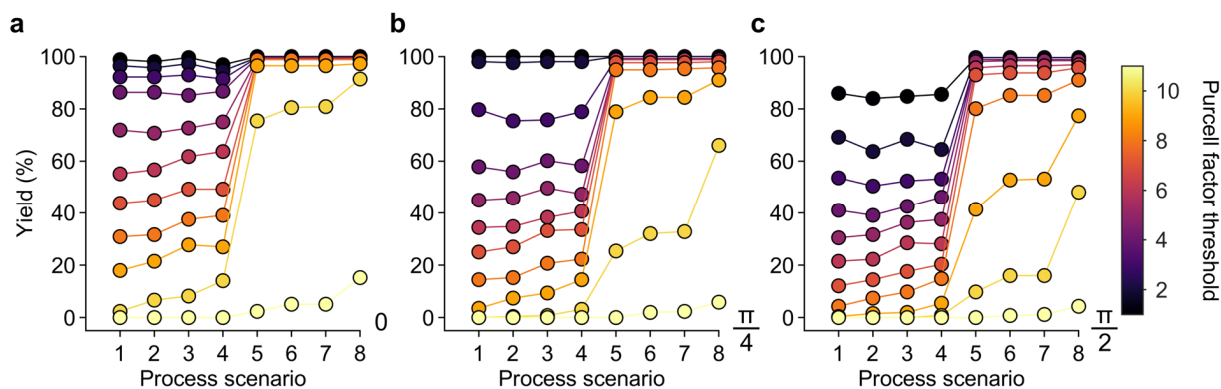

**Figure 8.** Yield improvement due to accurate integration. (**a-c**) Plots showing theoretical yield for each process scenario, for dipole angles of (a) 0 rad, (b) π/4 rad, and (c) π/2 rad. The color scale indicates the minimum value of Purcell factor corresponding to accurate integration. Data are representative and correspond to the mean value of Purcell factor for each field position (Figure 7, Supplementary Figure S9). Lines connecting data points are for visual guidance. An alternate presentation of these data is in Supplementary Figure S10.

**CONCLUSION**

In this study, we establish a firm foundation of traceability of localization microscopy to the SI. This new capability is fundamental as localization microscopy matures, requiring not only novel methods, but also reliable quantities for meaningful comparisons across studies and systems.

We begin with the first correlation of aperture arrays by atomic-force and optical microscopy, yielding a master standard for optical calibration. The localization data correlate to within a few nanometers (Supplementary Table S1), which is interesting considering that the aperture sidewalls are rough at a scale of tens of nanometers. These results open the door to traceable correlations of surface structure and optical signal, extending down toward the atomic scale, across a wide field.

Using this master standard, we establish the concept of an uncertainty field in localization microscopy, shifting the paradigm from the typical focus on precision to a broader understanding of accuracy[8] and further to traceability, with a critical limit due to scale factor calibration. The concern of scale factor uncertainty or error depends on localization precision and field area. For super-resolution imaging, a typical count of $10^3$ to $10^4$ signal photons per fluorophore image limits localization precision to a few tens of nanometers within a field area of a few square micrometers. In this context, scale effects can be negligible. However, a stable and intense emitter can yield $10^6$ signal photons per image, with localization precision extending into the picometer scale by averaging replicates (Table 2)[10,46]. In this context, scale effects, which increase with distance, can be dominant across virtually any field area, presenting a critical limitation to localization accuracy.

Our approach extends a new level of traceability to ordinary optical microscopes in the field, upgrading them to metrology systems with high throughput. In this way, we develop and apply critical-dimension localization microscopy to characterize multiple working standards, validating lithographic accuracy on average at the subnanometer scale and opening the door to statistical process control to enable reliance on lithography systems to set critical dimensions. Future studies could optimize the process and establish tighter control than our prediction interval of less than 3 nm to produce working standards with high quality and efficiency for dissemination, potentially without the need for individual characterization. Finally, our standards are integrable for *in situ* calibration of beam placement and emitter position[56].

We apply these new capabilities to comprehensively calibrate a localization microscope with a sample cryostat and custom optics. Our study elucidates subtle but critical aspects of calibrating scale factor and correcting complex distortion, meeting multiple challenges and presenting new opportunities for cryogenic microscopy. After calibration, and in combination with a reference thermometer, a cryogenic microscope could serve as a localization dilatometer to measure the coefficient of thermal expansion of microscale structures and devices, rather than bulk materials. Alternately, after calibration and in combination with a reference material, a cryogenic microscope could function as a localization thermometer at the sample position, rather than elsewhere in the sample cryostat. The coefficient of thermal expansion of gallium arsenide is potentially suitable for localization thermometry, motivating its traceable characterization as a reference material.

Building on our foundation of traceability, we introduce a comprehensive model of registration errors in a theoretical process of integrating quantum emitters and photonic structures—a foundation for high yield in quantum engineering. This key advance unlocks the full potential of the most common and most scalable method of widefield microscopy for such an integration process, in contrast to other relevant methods of scanning confocal and cathodoluminescence microscopy. Our theory demonstrates the possibility of greatly improving the distribution of Purcell factor across an ultrawide field and dramatically increasing yield, enabling a transition from demonstration devices to reliable processes. In turn, this advance will enable the statistical characterization of device performance for quantum information systems.

**DATA AVAILABILITY**
The data supporting the findings of this study are available from the corresponding author upon reasonable request.

**CODE AVAILABILITY**
The code supporting the findings of this study are available from the corresponding author upon reasonable request.

**ACKNOWLEDGEMENT**
The authors acknowledge John Kramar for an insightful review of an early draft of this manuscript, Muneesh Maheshwari for helpful comments on custom optics, and funding from the NIST Innovations in Measurement Science Program and the NIST Office of Reference Materials.

## AUTHOR CONTRIBUTIONS

SMS supervised the study. SMS and CRC conceived the study. CRC performed optical microscopy, data analysis, and Monte-Carlo simulations with contributions from SMS. ALP developed statistical models and performed statistical analysis. RGD performed atomic-force microscopy. AC, KS, and MID developed cryogenic optical microscopes and performed measurements with them. DAW fabricated pillar arrays. BRI fabricated aperture arrays. CRC and SMS wrote the manuscript with contributions from all authors.

## COMPETING INTERESTS

The authors declare no competing interests.

# Supplementary Information *for*

# Traceable localization enables accurate integration of quantum emitters and photonic structures with high yield

Craig R. Copeland,[1] Adam L. Pintar,[2] Ronald G. Dixson,[1] Ashish Chanana,[1] Kartik Srinivasan,[1,3] Daron A. Westly,[1] B. Robert Ilic,[1,4] Marcelo I. Davanco,[1] and Samuel M. Stavis[1,*]

[1]Microsystems and Nanotechnology Division, [2]Statistical Engineering Division, [4]CNST NanoFab, National Institute of Standards and Technology, Gaithersburg, Maryland 20899, USA

[3]Joint Quantum Institute, NIST/University of Maryland, College Park, Maryland 20742, USA

[*]samuel.stavis@nist.gov

**INDEX**

**Table S1. Aperture pair distances**

| Aperture pair $i$ | Distance by AFM $D_{ij}^{AFM}$ (nm) | | | | | Distance by OM $D_{ij}^{OM}$ (nm) | | Deviation $D_{i\cdot}^{OM} - D_{i\cdot}^{AFM}$ (nm) |
|---|---|---|---|---|---|---|---|---|
| | $D_{i1}^{AFM}$ | $D_{i2}^{AFM}$ | $D_{i3}^{AFM}$ | $D_{i\cdot}^{AFM}$ | $\sigma_{D_{ij}^{AFM}}$ | $D_{i\cdot}^{OM}$ | $\sigma_{D_{ij}^{OM}}$ | |
| 1 | 4999.89 | 4999.18 | 5000.70 | 4999.93 | 0.76 | 5000.47 | 0.31 | 0.54 |
| 2 | 4999.21 | 5002.24 | 5001.58 | 5001.01 | 1.60 | 5000.08 | 0.52 | -0.93 |
| 3 | 5001.08 | 5001.00 | 4998.91 | 5000.33 | 1.23 | 5002.40 | 0.57 | 2.07 |
| 4 | 5000.49 | – | 5001.05 | 5000.77 | 0.40 | 5000.22 | 0.58 | -0.55 |
| 5 | 5001.82 | 5003.17 | 4998.66 | 5001.22 | 2.31 | 5000.03 | 0.59 | -1.18 |
| 6 | 4999.28 | 4998.04 | 5001.28 | 4999.53 | 1.63 | 5000.34 | 0.58 | 0.81 |
| 7 | 5001.58 | 5000.25 | 5000.61 | 5000.82 | 0.69 | 5001.94 | 0.59 | 1.12 |
| 8 | 4999.57 | 5000.87 | 5000.02 | 5000.15 | 0.66 | 4999.03 | 0.58 | -1.13 |
| 9 | 5001.71 | 5001.50 | 5003.28 | 5002.16 | 0.97 | 5002.44 | 0.55 | 0.29 |
| 10 | 5001.45 | 5001.78 | 4998.40 | 5000.55 | 1.86 | 5000.22 | 0.50 | -0.32 |
| 11 | 5002.85 | 5000.61 | 5000.98 | 5001.48 | 1.20 | 5000.18 | 0.31 | -1.30 |
| 12 | 5002.97 | 5001.81 | 5002.76 | 5002.51 | 0.62 | 5004.01 | 0.34 | 1.50 |
| 13 | 4999.00 | 4999.16 | 4999.07 | 4999.08 | 0.08 | 4998.27 | 0.52 | -0.81 |
| 14 | 5001.27 | 5001.53 | 5001.48 | 5001.43 | 0.14 | 5001.26 | 0.56 | -0.17 |
| 15 | 5000.31 | 5000.52 | 4999.87 | 5000.24 | 0.33 | 5000.55 | 0.56 | 0.32 |
| 16 | 5000.37 | 5000.46 | 5000.34 | 5000.39 | 0.06 | 4999.59 | 0.55 | -0.80 |
| 17 | 5002.16 | 5002.20 | 5001.85 | 5002.07 | 0.19 | 5002.41 | 0.58 | 0.34 |
| 18 | 4999.57 | 4999.66 | 4999.96 | 4999.73 | 0.20 | 4999.71 | 0.58 | -0.02 |
| 19 | 5001.71 | 5002.04 | 5001.73 | 5001.83 | 0.19 | 5001.11 | 0.58 | -0.71 |
| 20 | 4999.26 | 4998.60 | 4998.79 | 4998.88 | 0.34 | 5000.79 | 0.57 | 1.91 |
| 21 | 5000.01 | 4998.92 | 4999.30 | 4999.41 | 0.55 | 4999.81 | 0.53 | 0.34 |
| 22 | 5001.79 | 5002.50 | 5001.93 | 5002.07 | 0.38 | 5002.23 | 0.36 | 0.16 |

Aperture pairs 1 to 11 correspond to AFM axis 1 and OM y axis.
Aperture pairs 12 to 22 correspond to AFM axis 2 and OM x axis.
Distance by OM results from 1000 replicate measurements.
The mean deviation is 0.01 nm ± 0.24 nm.

**Note S1. Uncertainty evaluation and expression**

We evaluate uncertainty using multiple methods that are fit for purposes that vary throughout our study, ranging from Gaussian error propagation to Bayesian statistical models. Accordingly, we cite five references, including three formal guidelines[1-3] that are in use at the National Institute of Standards and Technology (NIST), and two research publications from NIST[4,5]. Moreover, we aim to express uncertainty in a way that is clear to microscopists, metrologists, and statisticians, meriting further discussion.

An important issue to consider is an ambiguity of terminology. The original guidelines for evaluating and expressing the uncertainty of NIST measurement results[1] refer to a standard uncertainty $u$ that is an estimated standard deviation. However, depending on the context, $u$ can correspond either to the standard deviation $\sigma$ of a distribution, or to the standard error $\varepsilon$ of the mean $\mu$ of a distribution, $\varepsilon_\mu = \sigma/\sqrt{n}$, or even to the standard error $\varepsilon$ of the standard deviation $\sigma$ of a distribution, $\varepsilon_\sigma$, as in Supplementary Reference 6. The original guidelines then refer to an expanded uncertainty, $U$, that is a multiple of $u$, defining a symmetric interval of width $\pm U$ within which the value of the measurand is confidently believed to lie. The original guidelines further state that, for certain conditions, such an interval is a confidence interval. However, this statement does not resolve the ambiguity, because if $u$ corresponds to $\varepsilon_\mu$, then the term confidence interval can be appropriate, but if $u$ corresponds to $\sigma$, then the term prediction interval is nearer to correct. Supplementary Reference 7 illustrates many types of statistical, or probabilistic per the reference, intervals in use in metrology.

Another important issue to consider is the distinction between Type A and Type B evaluations of uncertainty, respectively involving statistical and other methods. The original guidelines state that the term confidence interval, in association with its frequentist statistical meaning, is appropriate only for Type A evaluations of uncertainty. In the case of Type B evaluations of uncertainty, none of the terms that are typical of classical or frequentist intervals are appropriate. Bayesian intervals may still be appropriate in the presence of Type B evaluations of uncertainty.

For these reasons and for clarity, we report all uncertainties in our study as 68 % coverage intervals. The term coverage interval is common in metrology, and we use it generically for any interval with a corresponding probability, although this term has its own meaning in some contexts. For example, in Supplementary References 7 and 8, this term clarifies the interpretation of the corresponding probability. Our coverage intervals often result primarily from Type A evaluations of standard errors of the means of distributions. In isolation, these values of $\varepsilon_\mu$ correspond to confidence intervals. However, the necessity of Type B evaluations of other components of uncertainty motivates our use of the term coverage interval.

For example, replicate measurements by optical microscopy of an aperture image result in a confidence interval of its apparent mean position in units of pixels, which is equal to the empirical localization precision in units of pixels over the square root of the number of replicate localization measurements. Conversion of the mean position and confidence interval to units of nanometers requires multiplication by a scale factor in units of nanometers per pixel. The scale factor uncertainty involves Type B evaluations of some components of uncertainty. Accordingly, the confidence interval becomes a coverage interval upon conversion of units from pixels to nanometers. In one case that we identify, the relevant coverage interval is a prediction interval for single samples.

**Table S2. Microscopy scale factor uncertainty**

| Uncertainty component | Relative value |
|---|---|
| Uncertainty of transfer standard for calibration of AFM scale factor | $6.75 \times 10^{-5}$ |
| Variability of replicate calibrations of AFM scale factor | $7.90 \times 10^{-5}$ |
| Variability of pitch estimate from AFM measurements | $2.60 \times 10^{-5}$ |
| Total uncertainty of scale factor for calibration and application of OM | $\sigma_S = 1.07 \times 10^{-4}$ |

Scale factor uncertainties involve both Type A and Type B evaluations of several components of uncertainty. We approximate the total uncertainty of scale factor for calibration and application of OM as a 68 % coverage interval for a normal distribution.

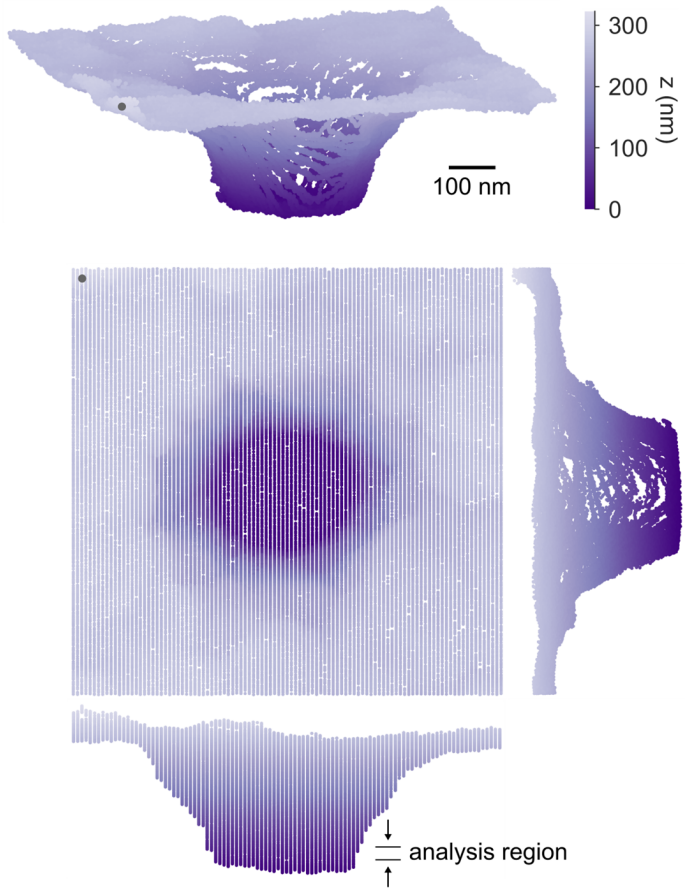

**Figure S1.** Representative aperture sidewalls. Atomic-force micrograph showing a representative aperture from different perspectives. Gray dots indicate the same corner of the micrograph. The data structure results from scan axis 2 of the atomic-force microscope. Black arrows indicate the sidewall region that we select for localization analysis.

**Note S2. Statistical models of distance**

We develop two statistical models to analyze aperture pair distances. The models are complementary and yield consistent results. Both models account for variability that is observable within each set of measurement results from the two microscopy methods. The fixed-effect linear model accounts for additional variability that is unobservable within each set of measurement results, due to localization errors in both atomic-force microscopy and in optical microscopy. We can independently estimate the former errors and solve for the latter errors in a comparison of the two sets of measurement results. The autoregressive–moving-average model incorporates aperture placement as another component of uncertainty, yielding slightly larger 68 % coverage intervals. Neither statistical model accounts for scale factor uncertainty, which is common to both models but is unobservable in any aspect of our measurements (Supplementary Table S1, Supplementary Table S3). We propagate the scale factor uncertainty subsequently (Supplementary Note 3).

*Fixed-effect linear model*

For atomic-force microscopy, the distance from replicate measurement $j$ of aperture pair $i$ is $D_{ij}^{AFM} = \Delta_i + d_{ij}^{AFM} + \delta_i^{AFM}$, where $\Delta_i$ is the true distance between aperture pair $i$, $d_{ij}^{AFM}$ is a random error that is observable between replicate measurements of aperture pair $i$, and $\delta_i^{AFM}$ is a random error that is unobservable between replicate measurements of aperture pair $i$. It is necessary to allow each aperture pair to have its own true distance $\Delta$, because for two adjacent pairs, if the second aperture is closer to the first aperture, then it is further from the third aperture. Such a correlation suggests an alternate analysis by an autoregressive–moving–average model, as we describe subsequently. For each axis, the mean distance between aperture pairs is $\frac{1}{11}\sum_{i=1}^{11} \Delta_i + d_{i\cdot}^{AFM} + \delta_i^{AFM}$. For the available data, estimation of both $\Delta_i$ and $\delta_i^{AFM}$ is impossible, but estimation of their sum is possible. Additional data allows estimation of the magnitudes of the $\delta_i^{AFM}$. We take as an estimate of the array pitch the average of the mean distances from each axis, neglecting the possibility of a correlation of those mean distances because they share a corner aperture. Neglecting this possible correlation is a tolerable simplification because, as we describe in the next section, higher variability of replicate measurements obscures correlations between adjacent pairs of apertures along axis 1 (Supplementary Figure S2).

For optical microscopy, the distance from replicate measurement $j$ of aperture pair $i$ is $D_{ij}^{OM} = \Delta_i + d_{ij}^{OM} + \delta_i^{OM}$, where $\Delta_i$ is the same true distance as in the fixed-effect linear model for atomic-force microscopy, $d_{ij}^{OM}$ is a random error that is observable between replicate measurements of aperture pair $i$, and $\delta_i^{OM}$ is a random error that is unobservable in replicate measurements of aperture pair $i$.

Distance deviations between the two methods are $D_{i\cdot}^{OM} - D_{i\cdot}^{AFM} = \delta_i^{OM} + d_{i\cdot}^{OM} - \delta_i^{AFM} - d_{i\cdot}^{AFM}$, where the dot for $j$ in the subscript denotes averaging over replicates. We denote the variance of $D_{i\cdot}^{OM} - D_{i\cdot}^{AFM}$ by $\sigma_{D^{OM}-D^{AFM}}^2$, the variance of $\delta_i^{AFM}$ by $\sigma_{\delta^{AFM}}^2$, the variance of $\delta_i^{OM}$ by $\sigma_{\delta^{OM}}^2$, the variance of $d_{i\cdot}^{AFM}$ by $\sigma_{d^{AFM}}^2/n_{D^{AFM}}$, and the variance of $d_{i\cdot}^{OM}$ by $\sigma_{d^{OM}}^2/n_{D^{OM}}$ (Table 1).

*Autoregressive–moving-average model*

In the fixed-effect linear model, we account for the fact that adjacent pairs of apertures share one aperture by assuming distinct distances for each pair. However, we might also assume a single mean distance for all pairs of apertures along an axis, but allow for autocorrelation between them. We expect a negative autocorrelation between distance measurements of adjacent pairs, and we use an autoregressive–moving-average model[9] to allow and account for this effect. The model applies to averages across replicates but does not apply to the individual replicates (Supplementary Figure S2). The typical application of such a model is to a time series, whereas our measurement results have a spatial, rather than a temporal, distribution. Nonetheless, along each scan axis of the atomic force microscope, the spatial distribution is one-dimensional, neglecting off-axis effects of fabrication precision that are small, so that the model is still applicable. A negative autocorrelation is clearly present for axis 2, but is only faintly evident for axis 1 (Supplementary Figure S2). These results are consistent with the respectively lower and higher variability of replicate measurements along the scan axes (Supplementary Figure S2). Zero autocorrelation is a limiting case for autoregressive–moving-average models, so the same statistical methodology is applicable for both axes.

A challenge in applying the autoregressive–moving-average family of models is choosing a member. To this end, we apply the methodology of Supplementary Reference 10, using common software for statistical analysis[11]. For axis 1, this methodology selects a model that does not include autocorrelation, that is, a model that takes the distances to be statistically independent. For axis 2, the methodology selects a moving average model of order 1. To test these results, we simulate hypothetical values from the model selections for comparison to the measurement results (Supplementary Figure S2). The simulation and measurement results are in good agreement, implying that the selections of the methodology are fit for this purpose.

For axis 1, the mean distance is 5000.72 nm $\pm$ 0.21 nm, for axis 2 it is 5000.57 nm $\pm$ 0.19 nm, and for all 22 pairs, it is 5000.64 nm $\pm$ 0.14 nm. The estimate of diagonal distance is 77791.8 nm $\pm$ 2.5 nm. These results are in good agreement with the results of the fixed-effect linear model in the main text. In most cases, the 68 % coverage intervals from the autoregressive-moving-average models are slightly larger. This is because the autoregressive models incorporate aperture placement as an additional component of uncertainty into their assessments. The additional capability results from accounting for differences in distances through correlation effects of second order, instead of mean effects of first order.

**Table S3. Distance uncertainty observability**

| Extent of observability | Component of uncertainty or variability |
|---|---|
| Observable between replicate measurements by AFM or OM | $\sigma^2_{d^{AFM}}, \sigma^2_{d^{OM}}$ |
| Observable between correlative measurements by AFM and OM | $\sigma^2_{\delta^{AFM}}, \sigma^2_{\delta^{OM}}$ |
| Unobservable between correlative measurements by AFM and OM | $\sigma_S$ |

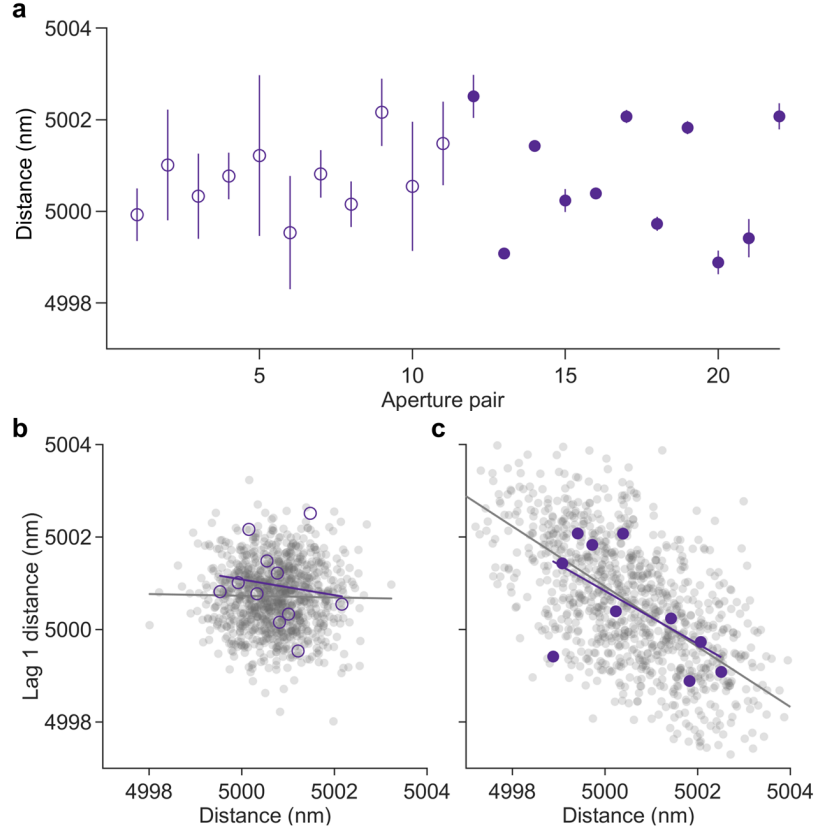

**Figure S2.** Autocorrelation analyses. (a) Plot showing distance measurements along (hollow circles) axis 1 and (solid circles) axis 2 of the atomic force microscope. Uncertainties are 68 % coverage intervals from replicate measurements. (b, c) Plots showing autocorrelation analyses of distance measurements along (b) axis 1 and (c) axis 2. Vertical axes show lag 1 distances. Horizontal axes show distances between aperture pairs. Purple data points are measurement results. Purple lines are least-squares fits to measurement results. Gray data points are simulation results from the model selection. Gray lines are least-squares fits to simulation results. (c) A few gray data points lie outside of the plot range, which maximizes clarity. The variability of the data that is observable is one component of the total uncertainty. We subsequently propagate uncertainty of the scale factor.

**Table S4. Distance deviation variances**

| AFM axis | OM axis | $\sigma^2_{d^{AFM}}$ (nm$^2$) | $\sigma^2_{d^{AFM}}/n_{D^{AFM}}$ (nm$^2$) | $\sigma^2_{d^{OM}}$ (nm$^2$) | $\sigma^2_{d^{OM}}/n_{D^{OM}}$ (nm$^2$) | $\sigma^2_{D^{OM}-D^{AFM}}$ (nm$^2$) |
|---|---|---|---|---|---|---|
| 1 | y | 1.86 | 0.62 | 0.28 | 0.00028 | 1.22 |
| 2 | x | 0.11 | 0.04 | 0.28 | 0.00028 | 0.77 |

Variances of replicate measurements pool over aperture pairs for both microscopy methods.

**Note S3. Distance uncertainty for optical microscopy**

The Euclidean distance $D^{OM}$ between two points in two lateral dimensions depends on the positions of the points in the $x$ and $y$ directions. In the main text, Eq. (4) describes the total uncertainty of position in a single lateral dimension for the case that one of the points is a reference position without localization uncertainty, so that the variance of $D_x^{OM}$ reduces by a factor of two, $\sigma_{D_x^{OM}}^2/2$. The uncertainty is then,

$$u_{D_x^{OM}} = u_{D_x^{OM}}^{\text{I}} = \sqrt{\left(\sigma_{D_x^{OM}}/\sqrt{2}\right)^2 + \left(D_x^{OM} \times \sigma_S\right)^2}, \tag{S1}$$

where $\sigma_S$ is the relative uncertainty of the scale factor. The y direction has an analogous expression. We denote whether one or two points are the result of localization analysis with corresponding uncertainty by superscript Roman numerals I and II, respectively. We omit this notation from the main text for concision. The division of $\sigma_{D_x^{OM}}$ by $\sqrt{2}$ is due to our quantification of $\sigma_{d^{OM}}$ and $\sigma_{\delta^{OM}}$ by measurements of distance between two points (Table 2), assuming equal localization uncertainty for each point.

We derive a general expression for the uncertainty of $D^{OM}$, for the case that the positions of both points are the result of localization analysis, $u_{D^{OM}}^{\text{II}}$ (Table 1). In this case, the standard deviation of $D_x^{OM}$ is $\sigma_{D_x^{OM}}$, as the positions of both points are subject to localization uncertainty. The values of $D_x^{OM}$ and $D_y^{OM}$ are in units of nanometers after multiplication of the scale factor for optical microscopy. To limit the complexity of our notation, we express measurement equations that account for the relative uncertainty of scale factor $\sigma_S$ through a multiplicative scale factor $S = 1$, so that the distances in the $x$ and $y$ directions are the products $SD_x^{OM}$ and $SD_y^{OM}$, respectively. An equivalent derivation, requiring additional notation beyond that of Table 1, could instead use measurements of distance in units of pixels, so that the scale factor and scale factor uncertainty would be both in units of nanometers per pixel. By linearizing $SD_x^{OM}$ and calculating the standard deviation of the resulting linear approximation, we determine an approximate uncertainty,

$$u_{D_x^{OM}}^{\text{II}} \approx \sqrt{\sigma_{D_x^{OM}}^2 + \left(D_x^{OM} \times \sigma_S\right)^2}, \tag{S2}$$

which is Eq. (4) and Eq. (S1) without the factor $1/\sqrt{2}$. The Euclidean distance is then,

$$D^{OM} \approx \sqrt{\left(D_x^{OM}\right)^2 + \left(D_y^{OM}\right)^2}. \tag{S3}$$

Again, by linearizing $D^{OM}$ and calculating the standard deviation of the resulting linear approximation, we arrive at an approximate expression for the uncertainty of $D^{OM}$,

$$u_{D^{OM}}^{\text{II}} \approx \sqrt{\frac{\left(D_x^{OM}\right)^2 \sigma_{D_x^{OM}}^2 + \left(D_y^{OM}\right)^2 \sigma_{D_y^{OM}}^2}{\left(D_x^{OM}\right)^2 + \left(D_y^{OM}\right)^2} + \left(\left(D_x^{OM}\right)^2 + \left(D_y^{OM}\right)^2\right)\sigma_S^2}. \tag{S4}$$

Eq. (S2) and Eq. (S4) are similar. In Eq. (S2), the first term is the square of the uncertainty for a single lateral dimension. In Eq. (S4), the first term is a weighted average of the square of the distance uncertainties for both lateral dimensions $x$ and $y$. In both equations, the second term is the product of the square of the relative scale factor uncertainty and the square of the distance between the points.

**Table S5. Master and working standards**

| Standard arrays | | | | | | | | | |
|---|---|---|---|---|---|---|---|---|---|
| Standard type | Master | Working | Working | Working | Working | Working | Working | Working | Working |
| Feature type | Aperture | Aperture | Aperture | Aperture | Aperture | Aperture | Pillar | Pillar | Pillar |
| Array number | 1 | 2 | 3 | 4 | 5 | 6 | 7 | 8 | 9 |
| **Lithography parameters** | | | | | | | | | |
| Lithography system 1 or 2 | 1 | 1 | 2 | 2 | 2 | 2 | 2 | 2 | 2 |
| Electron-beam distortion correction | yes | no | yes | yes | yes | yes | yes | yes | yes |
| Electron-beam current (nA) | 1.000 | 1.000 | 1.000 | 1.000 | 0.125 | 0.200 | 1.000 | 1.000 | 1.000 |
| Exposure passes | 1 | 1 | 1 | 1 | 8 | 1 | 1 | 1 | 1 |
| **Microscopy results** | | | | | | | | | |
| Mean pitch (nm) | 5000.71 | 4999.90 | 5000.30 | 5001.45 | 4997.30 | 4996.44 | 5001.64 | 5001.60 | 5001.88 |
| Standard error of the mean (nm) | 0.13 | 0.03 | 0.01 | 0.03 | 0.08 | 0.08 | 0.07 | 0.07 | 0.07 |
| Degrees of freedom | 43 | 2 | 4 | 5 | 2 | 2 | 8 | 8 | 8 |
| Traceable uncertainty (nm) | 0.54 | 0.54 | 0.54 | 0.54 | 0.54 | 0.54 | 0.54 | 0.54 | 0.54 |

Lithography system 1 has a write field of 62.5 µm by 62.5 µm and a beam placement specification of 0.125 nm.

Lithography system 2 has a write field of 1 mm by 1 mm and a beam placement specification of 2 nm.

Aperture arrays 1 through 6 are independent whereas pillar arrays 7 through 9 are effectively replicates.

**Note S4. Lithographic pitch accuracy**

To test the lithographic pitch, we apply critical-dimension localization microscopy to measure the master aperture array, along with five additional aperture arrays and three pillar arrays that serve as working standards. All nine arrays have a nominal pitch of 5000 nm. In the electron-beam lithography process, we vary the current of the electron beam and the number of exposure passes while keeping the total dose constant (Supplementary Table S5). Further details of this process are in Supplemental Reference 12.

Array 2 consists of one substrate with three different arrays of 16 apertures by 16 apertures that we image, whereas arrays 3 through 6 are single arrays with lateral extents of 300 µm from which we image distinct subsets of 16 apertures by 16 apertures. We localize the apertures in each image. A similarity transformation between the aperture positions and those of the master aperture array determines a multiplicative scale factor relating the traceable pitch of the master standard to the pitch of each working standard (Supplementary Table S5). We perform similar measurements for pillar arrays (Supplementary Note S5).

A statistical meta-analysis[4] characterizes the variability of pitch across the arrays, taking as inputs the mean value of pitch and its standard error for each aperture array, and the mean and standard error across the three replicate pillar arrays (Supplementary Table S5). Our selections of prior distributions for this Bayesian statistical analysis are an improper flat distribution for the mean pitch, and truncated Cauchy distributions with a scale parameter of five for the standard deviation of pitch between arrays and the standard deviation of pitch between measurements of single arrays. This consensus builder determines a dark uncertainty of 2.70 nm, which we interpret as a measure of pitch variability due to variation of the lithography process, and a consensus mean pitch with a coverage interval that we expand by propagating the total uncertainty of scale factor (Supplementary Table S2) by Monte-Carlo methods, yielding a traceable value of 4999.80 nm ± 0.98 nm. We apply the uncertainty of scale factor after the consensus analysis, as this uncertainty applies in a common mode and would otherwise obscure the estimate of dark uncertainty. Finally, this analysis yields a prediction interval for the pitch of additional aperture arrays ranging from 4997.42 nm to 5002.13 nm. We interpret this prediction interval as a conservative estimate of the reliability of producing working standards, due to our deliberate variation of lithographic process parameters and device geometry and our use of conservative priors in the Bayesian analyses (Supplementary Table S9).

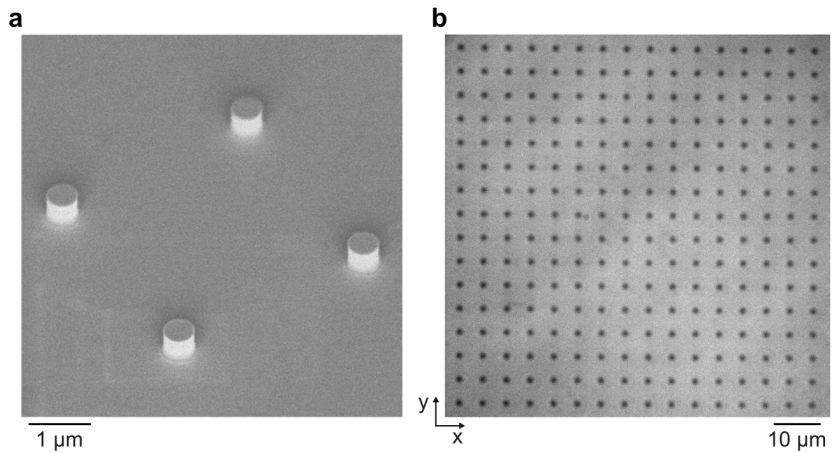

**Figure S3.** Silicon pillar arrays. (**a**) Scanning-electron micrograph from a perspective angle and (**b**) brightfield optical micrograph from the top down showing a representative pillar array in a silicon (100) substrate. The micrograph in (b) shows the subset of the array that we use for calibrating the cryogenic microscope.

**Note S5. Calibration of a cryogenic microscope**

The cryogenic microscope consists of a light-emitting diode with a nominal wavelength of 940 nm, an objective lens with a nominal magnification of 100× and a numerical aperture of 0.75, and a tube lens with a nominal focal length of 200 mm. The objective lens is within the cryogenic chamber of a closed-cycle helium cryostat, imaging the sample through a fused silica window with a thickness of 580 µm and an anti-reflective coating for a wavelength range of 400 nm to 1100 nm. The cryostat maintains the sample at a nominal temperature of 1.81 K. A three-axis piezoelectric actuator with a resolution of approximately 200 nm controls the sample position. An electron-multiplying charged-coupled device camera with an on-chip pixel size of 13 µm records brightfield reflection images with an integration time of 1 s.

For this microscope system in this imaging mode, we find that aperture arrays provide insufficient contrast for localization analysis. As well, the uncertain coefficient of thermal expansion of the silica substrate and any effect of the metallic multilayer that forms the aperture arrays could raise questions about the cryogenic applicability of reference data from measurements at room temperature. To address these issues, we fabricate arrays of pillars in silicon (100), having a nominal diameter of 500 nm, an approximate height of 500 nm, and a nominal pitch of 5000 nm (Supplementary Figure S3). We find that these structures provide sufficient contrast for imaging and localization. Importantly, modern reference data of extremely high quality[13] for the coefficient of thermal expansion of silicon (100) enables accurate estimates of the net change of array pitch at cryogenic temperatures.

We apply critical-dimension localization microscopy to measure the pitch of silicon pillar arrays at room temperature, using a conventional optical microscope that is similar to the one we use for correlative measurements of aperture arrays in the main text[12], but with a different objective lens having a magnification of 100× and a numerical aperture of 0.75. We image pillar arrays and an aperture array standard having a traceable value of pitch (Supplementary Table S5, working array 4) in reflection near best focus, and determine the positions of each feature by localization with a Gaussian approximation. In reflection, both pillars and apertures produce dark images on a light background. Interestingly, the resulting signal dips are amenable to localization analysis by a Gaussian approximation, but with inverse contrast relative to the common use of Gaussian approximations to localize signal peaks from bright images on a dark background. We determine the pitch of the pillar array relative to the aperture array by the scale factor of a similarity transformation between the positions in each array. Prior to determining the transformation, we develop correction functions of Zernike polynomials that have a unique number of consecutive terms, or Noll orders, for each array type for this microscope. Functions with the optimal number of terms capture as much of the apparent position errors from optical aberrations as possible, while ignoring apparent position errors from random effects such as photon shot noise, and actual position errors from the fabrication process. This optimization process consists of fitting correction models with an increasing number of terms to the position data from a replicate array of one type, and then applying the models to correct the position data from a second replicate array of the same type. The model of optimal order provides the best correction for the second array. In this process, models with too many terms begin to include components of position error that are spatially random and particular to a micrograph of the first array. These unique errors result from both the precision of feature placement in the fabrication process and the effect of photon shot noise in a single micrograph, increasing the position errors that remain after correction of the second array. We find optimum performance using Zernike polynomials with a Noll order of 22 for the x direction for both array types, 19 for the y direction for aperture arrays, and 25 for the y direction for pillar arrays. Each pillar array is 20 pillars by 20 pillars, and we image nine distinct regions of the aperture array containing 20 apertures by 20 apertures. The calibration results are in Supplementary Table S6.

To calibrate the cryogenic microscope, we first determine the number of terms for the correction model following the same procedure, finding optimal performance from models that include Zernike polynomials up to a Noll order 63 for the x direction and 58 for the y direction. Comparison of the pillar positions in the array design to the experimental pillar positions following correction yields root-mean-square errors of position with mean values and standard deviations over the three arrays of 14.58 nm ± 1.25 nm in x and 16.27 nm ± 1.90 nm in y across an imaging field of 75 µm by 75 µm. We estimate the components of these position errors for a representative array, including errors from empirical localization precision and fabrication precision[12], in order to isolate the localization error of the microscope. We determine empirical localization precision from replicate measurements of single pillar positions. This Type A evaluation of localization uncertainty captures any effects of unintentional motion of the imaging system, in contrast to an alternate Type A evaluation of the residuals of the localization fit, which may be insensitive to these sources of error, leading to a potential underestimate of localization uncertainty[14]. We determine an upper bound of errors due to fabrication precision by the root-mean-square errors of position relative to the array design, after correction, from the measurements with the critical-dimension microscope. Accounting for these components of error enables calculation of the localization error of the cryogenic microscope as a lower bound of traceable localization uncertainty (Supplementary Table S7) (Figure 7b-c).

We additionally test interpolant models to calibrate the position errors of the cryogenic microscope. Unlike correction models of Zernike polynomials, the performance of interpolant models is impossible to measure by the errors that remain after applying the model to the data, since these errors are zero by definition. Instead, the application of interpolant models from one array to correct the position data from a different array allows the determination of resulting errors. This process results in additional apparent errors with respect to the values in Supplementary Table S7, due to experimental factors that produce differences between the two micrographs of the two arrays, such as variation in focus[12] and, for the cryogenic microscope in particular, variation in lateral position at the scale of one micrometer. In such tests, interpolant models perform slightly better than Zernike models (Supplementary Table S8). The interpolant models capture all of the position errors at the pillar locations, so these results indicate lateral variation of localization errors at scales below the array pitch, which become evident upon applying the correction model to a second array with a different lateral position in the imaging field.

**Table S6. Standards for cryogenic microscopy**

| Pillar array | Pitch at 293 K (nm) | Pitch at 1.8 K (nm) | Image pixel size at 1.8 K (nm) |
|---|---|---|---|
| 7 | 5001.64 ± 0.54 | 5000.56 ± 0.54 | 107.28 ± 0.01 |
| 8 | 5001.60 ± 0.54 | 5000.53 ± 0.54 | 107.30 ± 0.01 |
| 9 | 5001.88 ± 0.54 | 5000.80 ± 0.54 | 107.27 ± 0.01 |

Uncertainties of pitch include negligible contributions of 0.07 nm from measurement of array pitch at 293 K, and 0.00015 nm from estimation of the net contraction of silicon at 1.8 K.[13] Temperatures are approximate.

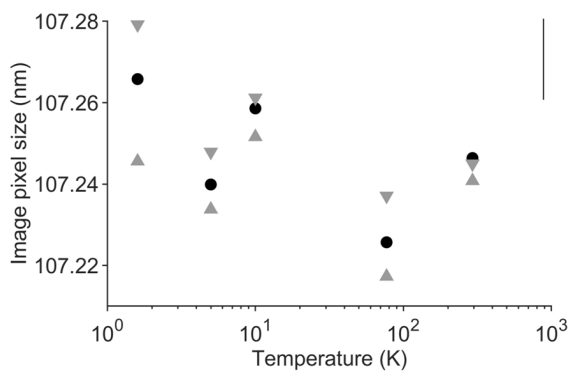

**Figure S4.** Magnification calibration of cryogenic microscope. Plot showing image pixel size as a function of temperature for a representative pillar array. Circle data markers are near best focus, triangles pointing down are below best focus, and triangles pointing up are above best focus. The lone bar at upper right indicates the total traceable uncertainty, which results predominantly from components of uncertainty that are common to all of these data. Components of uncertainty that are not common to all of these data are smaller than the data markers.

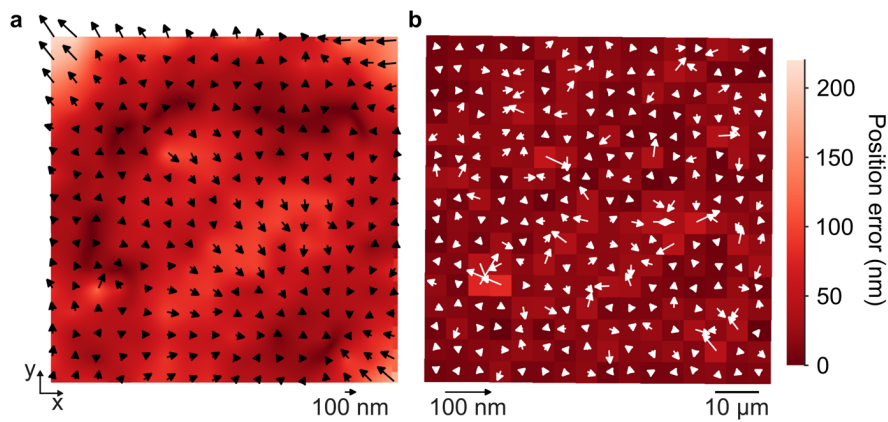

**Figure S5.** Distortion correction of cryogenic microscope at ambient temperature. Vector plots and color maps showing position errors (**a**) before and (**b**) after correction, for imaging at a temperature of approximately 293 K.

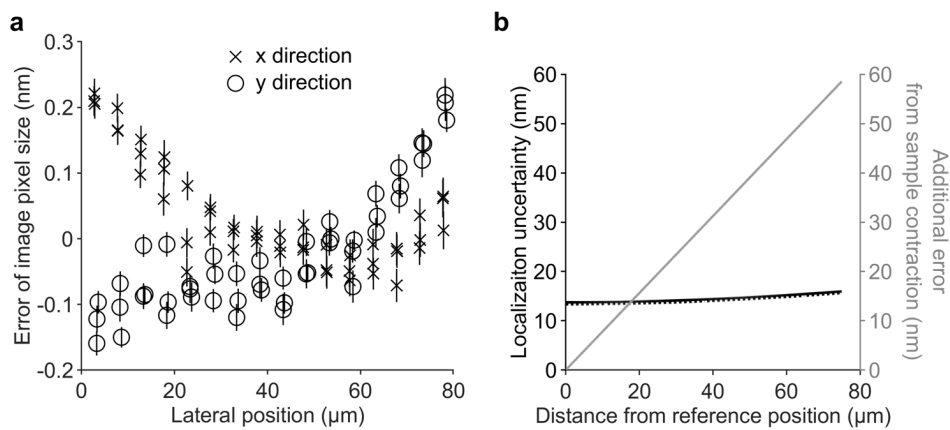

**Figure S6.** Potential errors in magnification calibration. (**a**) Plot showing error of image pixel size from calibration using as a scale reference the distance in the (crosses) x direction or (circles) y direction between two pillars on opposing sides of the field periphery, approximating the analysis of alignment markers of a frame type. Triplet data markers are from three different pillar arrays. (**b**) Plot showing the difference between (black) traceable localization uncertainty in the (solid) x and (dash) y directions, and (gray) localization error from the contraction of gallium arsenide[13] upon cooling to cryogenic temperatures, which results in a scale factor error of approximately $7.8 \times 10^{-4}$.

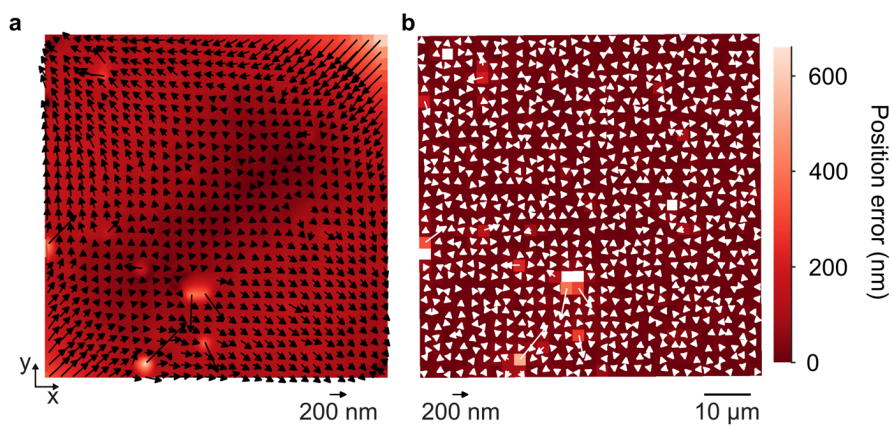

**Figure S7.** Distortion correction of additional cryogenic microscope at ambient temperature. Vector plots and color maps showing position errors (**a**) before and (**b**) after correction, for imaging at a temperature of approximately 293 K. This microscope is a modification of the system in reference 15 with a different objective lens with a numerical aperture of 0.7. This system is similar to the main cryogenic microscope in this study, except that there are no optical windows between the sample and the objective lens, which is in a sub-chamber of the sample cryostat.

**Table S7. Position errors in cryogenic localization microscopy**

| Pillar array | Total position error without correction (nm) | Total position error with correction (nm) | Localization precision (nm) | Fabrication precision (nm) | Localization error (nm) |
|---|---|---|---|---|---|
| 7 | 50, 54 | 14.7, 14.4 | 1.4, 1.4 | 3.5, 3.5 | 14.2, 13.9 |
| 8 | 45, 50 | 13.1, 15.5 | 1.2, 1.2 | 1.9, 2.5 | 12.9, 15.3 |
| 9 | 44, 48 | 16.2, 19.0 | 1.5, 1.4 | 1.8, 2.3 | 16.0, 18.8 |

Correction values are for a Zernike model.
All values are the root-mean-square over the array.
Commas separate values for the x and y directions.

**Table S8. Comparison of cryogenic correction models**

| Correction scheme | Total position error with Zernike correction (nm) | Total position error with interpolant correction (nm) |
|---|---|---|
| Correction of array 7 with the model from array 8 | 19.7, 18.1 | 18.0, 18.3 |
| Correction of array 8 with the model from array 7 | 18.0, 18.8 | 17.3, 18.5 |

Total position errors are root-mean-square values over all pillar positions of an array.
Commas separate values for the x and y directions.

**Note S6. Comprehensive model of registration errors**

Beginning with an array of $k$ positions $(x_{0,k}, y_{0,k})$ relative to the center of the imaging field and having a nominal pitch of 5000 nm, we simulate the effects of various sources of error as random values sampling probability distributions that we define for each error source (Supplementary Table S9). One exception is error due to distortion, for which we use the experimental values that correspond to each position in the field (Figure 6). The values of error sum and apply to each position. For the x direction,

$$x_k = \left[1 + \epsilon_{S_1} + \epsilon_{S_2} + \epsilon_{S_3}\right] \times x_{0,k} + \epsilon_{d,x}(x_{0,k}, y_{0,k}), \tag{S5}$$

where $\epsilon_{S_1}$, $\epsilon_{S_2}$, and $\epsilon_{S_3}$ are relative errors of scale factor from three sources, and $\epsilon_{d,x}$ is error from distortion, which depends on position. Although the characteristic definition of distortion is an effect of non-uniform magnification or scale factor, we separate this component of position error from uniform scale factor errors that do not explicitly depend on position. The y direction has an analogous expression.

We consider three sources of scale factor error that can be present in a magnification calibration. The first source of error is from using the nominal dimension of a standard, incurring an error that we define by the prediction interval of lithographic pitch from electron-beam lithography (Supplementary Note S4, Supplementary Table S9). The second source of error is from neglecting the net contraction of a gallium arsenide substrate that cools from approximately 293 K to approximately 1.8 K, incurring a relative error of approximately $7.8 \times 10^{-4}$ that we estimate from the data in Figure 2 of Supplemental Reference 16. Uncertainty of this quantity is absent from Supplemental Reference 16 and further evaluation is beyond the scope of the present study. The third source of error is from basing a magnification calibration on distance measurements between two alignment markers, sparsely sampling the effects of complex distortion throughout the imaging field. The resulting errors depend on the positions of the markers and can differ depending on whether the distance separating them is predominantly in the x or y direction (Supplementary Figure S6b). For each position in the field, we define a normal distribution for potential scale factor errors from this source, for alignment markers with separation predominantly in the y direction. Although elaboration is possible, we consider this representative analysis to be fit for our purpose, as it is preferable to altogether avoid this type of error in scale factor calibration by using an array standard.

We simulate the implementation of our calibration by removing scale factor errors and propagating scale factor uncertainty (Supplementary Table S6, Supplementary Table S9). We then replace the position errors from complex distortion (Figure 6a) with the position errors that remain following distortion correction (Figure 6b).

We complete our comprehensive model of lateral registration error $\epsilon_{reg,x}(x_k, y_k)$ in the integration process (Supplementary Figure S8) by adding to the position errors in Equation S5 additional errors from correlative lithography of photonic structures. Using our evaluation of electron-beam lithography in this study and a previous study[12], we include a fourth source of scale factor $\epsilon_{S_4}$ due to the prediction interval of lithographic pitch by electron-beam lithography (Supplementary Note S4), position error from the precision of feature placement $\epsilon_{fp,x}$, and position error from overlay or pattern alignment $\epsilon_{a,x}$ (Supplementary Table S9),

$$\epsilon_{reg,x}(x_k, y_k) = x_{0,k} - \left[(1 + \epsilon_{S_1} + \epsilon_{S_2} + \epsilon_{S_3} + \epsilon_{S_4}) \times x_{0,k} + \epsilon_{d,x}(x_{0,k}, y_{0,k}) + \epsilon_{fp,x} + \epsilon_{a,x}\right], \quad (S6)$$

with an analogous expression for the y direction. In some cases, the probability distribution that we use for a source of error involves representative experimental data with triplicate measurements (Supplementary Figure S6, Supplementary Table S6). In these cases, we use a Bayesian analysis assuming a normal distribution for the likelihood and with appropriate informative priors on the mean and standard deviation of the normal distribution[5], to estimate the distribution of errors (Supplementary Table S9). Distributions of Purcell factor (Supplementary Figure S8) for single positions in the field (Supplementary Figure S9) result from various sources of error and uncertainty (Supplementary Table S9). The results in Figure 7 of the main text are the mean values of such distributions at the nominal positions $(x_{0,k}, y_{0,k})$, ignoring negligible uncertainties from a large number of samples in the simulation.

**Table S9. Sources of registration error for the comprehensive model**

| Source | Data for estimation | Bayesian priors | Probability distribution |
|---|---|---|---|
| Scale standard | Note S4 | Note S4 | Posterior distribution, mean = 4999.8 nm, standard deviation = 2.70 nm |
| Cryogenic contraction | Supplementary Reference 16 | - | Approximate value without uncertainty, 3.9 nm / 5001.64 nm = 7.8×10$^{-4}$ |
| Sparse sampling of magnification | Figure S6a, circles | Mean: normal distribution, mean 0 nm, standard deviation 0.2 nm. Standard deviation: truncated normal distribution, mean 0 nm, scale parameter 0.05 nm | Posterior distribution, position dependent |
| Calibration uncertainty | Table S6 | Mean: normal distribution, mean 100 nm, standard deviation 10 nm. Standard deviation: truncated normal distribution, mean 0 nm, scale parameter 0.05 nm | Posterior distribution, mean = 107.28 nm standard deviation = 0.04 nm |
| Pattern alignment | Manufacturer specification | - | Normal distribution, mean = 0, standard deviation = 3.3 nm |
| Pattern placement | Supplementary Reference 12 | - | Normal distribution, mean = 0, standard deviation = 2.0 nm |
| Pattern scale | Note S4 | Note S4 | Posterior distribution, mean = 4999.8 nm, standard deviation = 2.70 nm |

Unitless values indicate relative errors.

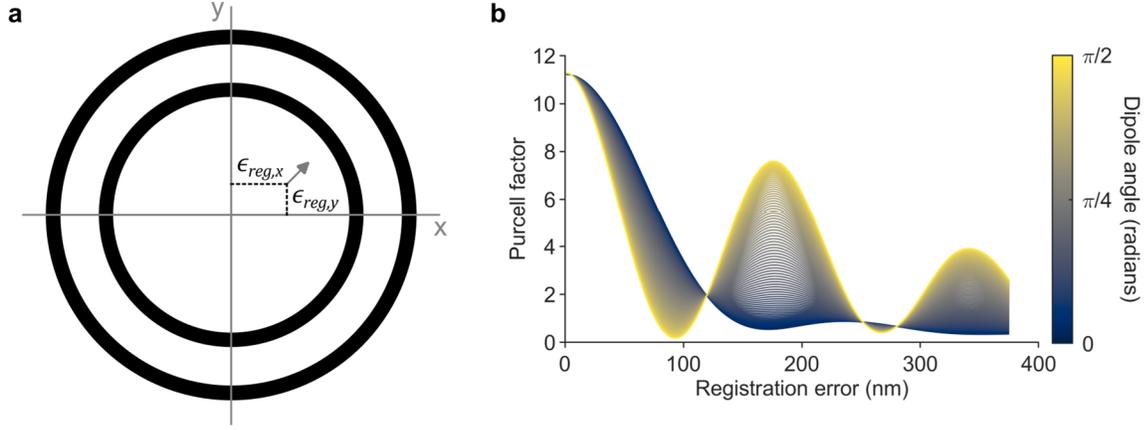

**Figure S8.** Purcell factor of a quantum dot inside a bullseye cavity. (**a**) Schematic showing a dipole emitter inside two trenches of a bullseye target cavity. The dipole emitter position has registration errors $\epsilon_{reg,x}$ and $\epsilon_{reg,y}$ relative to the cavity center. The dipole emitter angle is π/4 rad relative to the positive x direction, with a correspondence of color between (a) and (b). (**b**) Plot showing Purcell factor as a function of registration error for varying dipole angles relative to the positive x direction. To obtain these theoretical data, we interpolate the data from Supplementary Figure 6e of Supplementary Reference 14 for a wavelength of 948.02 nm. We neglect any uncertainty of Purcell factor.

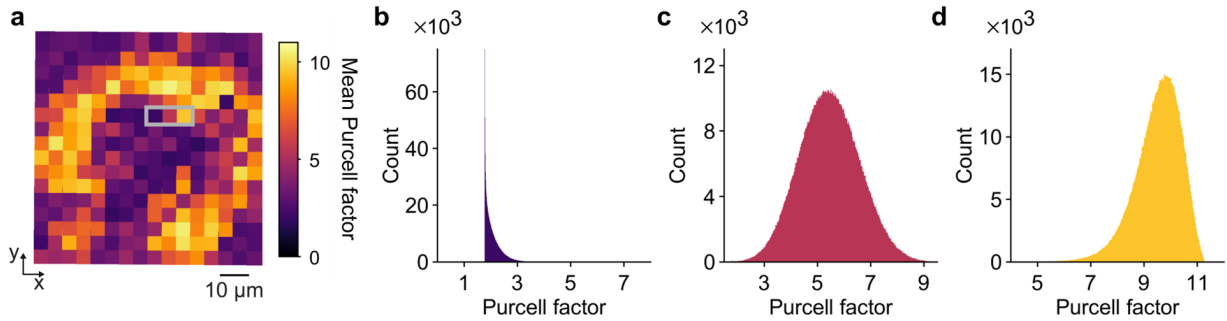

**Figure S9.** Purcell factor without distortion correction. (**a**) Plot showing mean Purcell factor across the imaging field for process scenario 4 and a dipole angle of 0 rad. The gray box indicates (**b-d**) histograms showing representative distributions of Purcell factor for three positions of adjacent pillars in (a). The colors correspond to the mean values.

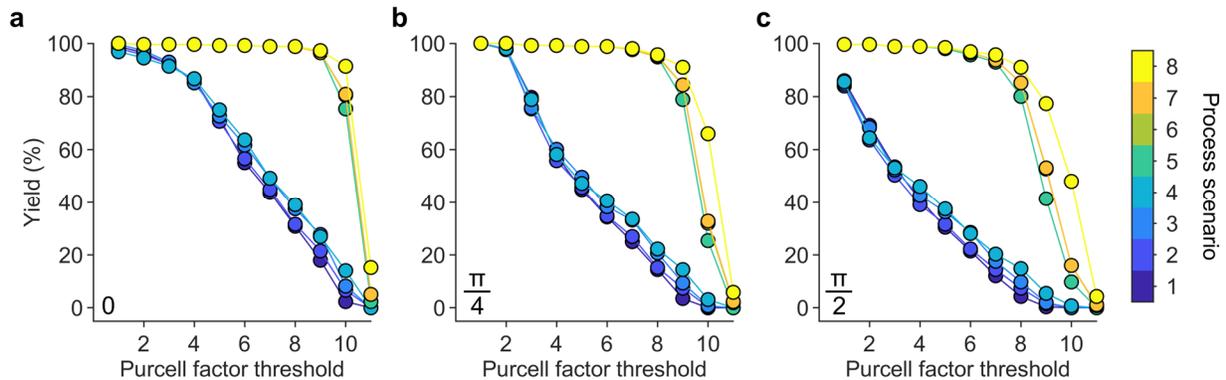

**Figure S10.** Yield reduction as a function of Purcell factor threshold. (**a-c**) Plots showing theoretical yield as a function of Purcell factor threshold, which is the minimum value of Purcell factor that is acceptable upon integration, for dipole angles of (a) 0 rad, (b) π/4 rad, and (c) π/2 rad. Colors indicate process scenarios. Data are representative and result from the mean value of Purcell factor for each field position (Figure 7, Supplementary Figure S9).

**Supplementary References**